%% file: main_arXiv_v2.tex
\documentclass[twocolumn,twocolappendix]{aastex63}
\usepackage{apjfonts} 

\usepackage[encapsulated]{CJK}  
\usepackage{xspace}  
\usepackage{makecell}
\usepackage{ucs}
\usepackage[utf8x]{inputenc}
\usepackage{amsmath}
\usepackage{multirow}  

\newcommand{\CASS}{\affiliation{Center for Astrophysics and Space Sciences, Department of Physics, University of California, San Diego, 9500 Gilman Drive, La Jolla, CA 92093, USA}}

\newcommand{\MPIE}{\affiliation{Max-Planck-Institut f\"ur extraterrestrische Physik, Giessenbachstra{\ss}e 1, D-85748 Garching, Germany}}
\newcommand{\MPIA}{\affiliation{
Max-Planck-Institut f\"ur Astronomie, K\"onigstuhl 17, D-69117 Heidelberg, Germany}}
\newcommand{\OSU}{\affiliation{Department of Astronomy, The Ohio State University, 4055 McPherson Laboratory, 140 West 18th Avenue, Columbus, OH 43210, USA}}
\newcommand{\UOA}{\affiliation{Department of Physics, University of Alberta, 4-183 CCIS, Edmonton, AB T6G 2E1, Canada}}
\newcommand{\ZAH}{\affiliation{Astronomisches Rechen-Institut, Zentrum f\"{u}r Astronomie der Universit\"{a}t Heidelberg, M\"{o}nchhofstra\ss e 12-14, D-69120 Heidelberg, Germany}}
\newcommand{\NRAO}{\affiliation{National Radio Astronomy Observatory, 520 Edgemont Road, Charlottesville, VA 22903, USA}}
\newcommand{\IRAM}{\affiliation{IRAM, 300 rue de la Piscine, F-38406 Saint Martin d'H\`eres, France}}
\newcommand{\LERMA}{\affiliation{LERMA, Observatoire de Paris, PSL Research University, CNRS, Sorbonne Universit\'es, F-75014 Paris, France}}

\newcommand{\logt}{{\rm log}\textsubscript{10}}
\newcommand{\dex}{{\rm dex}\xspace}
\renewcommand{\micron}{\ensuremath{\mu{\rm m}}\xspace}
\newcommand{\nm}{{\rm nm}\xspace}
\newcommand{\cm}{{\rm cm}\xspace}
\newcommand{\Mpc}{{\rm Mpc}\xspace}
\newcommand{\kpc}{{\rm kpc}\xspace}
\newcommand{\Myr}{{\rm Myr}\xspace}

\newcommand{\GHz}{{\rm GHz}\xspace}
\newcommand{\SigmaMassUnit}{{\rm M_\odot\,pc\textsuperscript{-2}}\xspace}
\newcommand{\SigmasfrUnit}{\ensuremath{{\rm M_\odot\,yr}^{-1}\,{\rm kpc}^{-2}}\xspace}
\newcommand{\IHIUnit}{{\rm Jy\,beam\textsuperscript{-1}\,km\,s\textsuperscript{-1}}\xspace}
\newcommand{\ICOUnit}{{\rm K\,km\,s\textsuperscript{-1}}\xspace}
\newcommand{\SEDUnit}{{\rm MJy\,sr\textsuperscript{-1}}\xspace}
\newcommand{\acoUnit}{\ensuremath{{\rm M_\odot\,pc^{-2}\,(K\,km\,s^{-1})^{-1}}}\xspace}
\newcommand{\XcoUnit}{\ensuremath{\rm cm^{-2}\,(K\,km\,s^{-2})^{-1}}\xspace}
\newcommand{\kappaUnit}{{\rm cm\textsuperscript{2}\,g\textsuperscript{-1}}\xspace}

\newcommand{\metal}{12+\logt({\rm O/H})\xspace}
\newcommand{\CO}{{\rm CO}\xspace}
\newcommand{\COonezero}{\ensuremath{{\CO\,J=1\to 0}}\xspace}
\newcommand{\COtwoone}{\ensuremath{{\CO\,J=2\to 1}}\xspace}
\newcommand{\HI}{\textsc{Hi}\xspace}
\newcommand{\HII}{\textsc{Hii}\xspace}
\newcommand{\HTWO}{{\rm H}\textsubscript{2}\xspace}
\newcommand{\Te}{{\rm T}\textsubscript{\rm e}\xspace}
\newcommand{\R}[1]{{\rm R}\textsubscript{#1}\xspace}
\newcommand{\Sigmad}{\ensuremath{\Sigma_d}\xspace}
\newcommand{\Sigmastar}{\ensuremath{\Sigma_\star}\xspace}
\newcommand{\Sigmasfr}{\ensuremath{\Sigma_{\rm SFR}}\xspace}
\newcommand{\Sigmagas}{\ensuremath{\Sigma_{\rm gas}}\xspace}
\newcommand{\Sigmametal}{\ensuremath{\Sigma_{\rm metal}}\xspace}
\newcommand{\Sigmaatom}{\ensuremath{\Sigma_{\rm atom}}\xspace}
\newcommand{\Sigmamol}{\ensuremath{\Sigma_{\rm mol}}\xspace}
\newcommand{\Sigmatot}{\ensuremath{\Sigma_{\rm Total}}\xspace}
\newcommand{\Bmaj}{{\rm B}\textsubscript{\rm maj}\xspace}
\newcommand{\Bmin}{{\rm B}\textsubscript{\rm min}\xspace}
\newcommand{\fgas}{{\rm f}\textsubscript{\rm gas}\xspace}
\newcommand{\fhtwo}{{\rm f}\textsubscript{\HTWO}\xspace}
\newcommand{\aco}{\ensuremath{\alpha_\CO}\xspace}
\newcommand{\acoMW}{\ensuremath{\alpha_\CO^{\rm MW}}\xspace}
\newcommand{\acoBolatto}{\ensuremath{\alpha_\CO^{\rm B13}}\xspace}
\newcommand{\acoSchruba}{\ensuremath{\alpha_\CO^{\rm S12}}\xspace}
\newcommand{\acoHunt}{\ensuremath{\alpha_\CO^{\rm H15}}\xspace}

\newcommand{\PDE}{\ensuremath{P_{\rm DE}}\xspace}
\newcommand{\PDEself}{\ensuremath{P_{\rm DE,\,self}}\xspace}
\newcommand{\PDEstar}{\ensuremath{P_{\rm DE,\,\star}}\xspace}

\definecolor{applegreen}{rgb}{0.55, 0.71, 0.0}

\shorttitle{Chiang et al.}
\shortauthors{Chiang et al.}

\begin{document}

\title{Resolving the Dust-to-Metals Ratio and CO-to-H\texorpdfstring{$_2$}{2} Conversion Factor in the Nearby Universe}

\correspondingauthor{I-Da Chiang}
\email{idchiang@ucsd.edu}

\author[0000-0003-2551-7148]{I-Da Chiang\begin{CJK*}{UTF8}{bkai}(江宜達)\end{CJK*}}\CASS
\author[0000-0002-4378-8534]{Karin M. Sandstrom}\CASS
\author[0000-0002-5235-5589]{J\'{e}r\'{e}my Chastenet}\CASS
\author[0000-0001-6405-0785]{Cinthya N. Herrera}\IRAM
\author[0000-0001-9605-780X]{Eric W. Koch}\UOA
\author[0000-0001-6551-3091]{Kathryn Kreckel}\MPIA\ZAH
\author[0000-0002-2545-1700]{Adam K. Leroy}\OSU
\author[0000-0003-3061-6546]{J\'{e}r\^{o}me Pety}\IRAM\LERMA
\author{Andreas Schruba}\MPIE
\author[0000-0003-4161-2639]{Dyas Utomo}\OSU\NRAO
\author[0000-0002-0012-2142]{Thomas Williams}\MPIA
\nocollaboration{11}

\begin{abstract}
\noindent We investigate the relationship between the dust-to-metals ratio (D/M) and the local interstellar medium environment at $\sim$2\,kpc resolution in five nearby galaxies: IC342, M31, M33, M101, and NGC628. A modified blackbody model with a broken power-law emissivity is used to model the dust emission from 100 to 500\,\micron observed by \textit{Herschel}. We utilize the metallicity gradient derived from auroral line measurements in \HII regions whenever possible. Both archival and new \CO rotational line and \HI 21\,cm maps are adopted to calculate gas surface density, including new wide field \CO and \HI maps for IC342 from IRAM and the VLA, respectively. We experiment with several prescriptions of \CO-to-\HTWO conversion factor, and compare the resulting D/M-metallicity and D/M-density correlations, both of which are expected to be non-negative from depletion studies. The D/M is sensitive to the choice of the conversion factor. The conversion factor prescriptions based on metallicity only yield too much molecular gas in the center of IC342 to obtain the expected correlations. Among the prescriptions tested, the one that yields the expected correlations depends on both metallicity and surface density. The 1-$\sigma$ range of the derived D/M spans 0.40${-}$0.58. Compared to chemical evolution models, our measurements suggest that the dust growth time scale is much shorter than the dust destruction time scale. The measured D/M is consistent with D/M in galaxy-integrated studies derived from infrared dust emission. Meanwhile, the measured D/M is systematically higher than the D/M derived from absorption, which likely indicates a systematic offset between the two methods.
\end{abstract}


\section{Introduction}\label{sec: Introduction}
Dust, the solid grains in the interstellar medium (ISM), plays an important role in shaping the interstellar radiation field and chemistry in the ISM. It absorbs or scatters a significant amount of starlight in galaxies \citep[e.g., 30\% suggested in][]{BERNSTEIN02}, and re-radiates in the infrared \citep[IR;][]{CALZETTI01, BUAT12}. Dust is important to the formation of molecular clouds because the surface of dust grains catalyze the formation of \HTWO \citep{GOULD63, DRAINE03, CAZAUX04, YAMASAWA11, GALLIANO18}, and dust grains can shield gas from the interstellar radiation field and help it cool to the temperature necessary for star formation \citep{KRUMHOLZ11, GLOVER12}.

In the diffuse ISM of the Milky Way (MW), around $20\%$ to $50\%$ of metals reside in dust grains according to elemental depletions \citep[$F_*=0$ to 1 in][]{JENKINS09}. This ratio of total metals locked in solid grains is called dust-to-metals mass ratio (D/M). D/M is important to ISM physics and offers constraints on dust chemical evolution. The equilibrium D/M represents a balance between dust formation and dust destruction. Among the dust evolution mechanisms, dust injection from the winds of asymptotic giant branch (AGB) stars, dust production from \mbox{Type~II} supernovae (SNe) and dust growth in the ISM are the major mechanisms that increase D/M, while dust destruction by SNe shock waves is the major mechanism that decreases D/M \citep{DWEK98,LISENFELD98,Draine09,ZHUKOVSKA16}.

Among these mechanisms, there is a broad consensus that dust growth in the ISM is a critical factor that sets D/M. Dust growth proceeds by accretion of gas-phase metals in the ISM onto existing dust grains, thus the dust growth rate should be positively correlated with metallicity and ISM gas density. Several models and simulations show that when the dust growth rate becomes higher than the dust destruction rate, D/M increases with metallicity and ISM gas density. As dust growth slows down as the gas-phase metals decrease, D/M becomes roughly constant \citep{DWEK98,HIRASHITA99, INOUE03_2003PASJ...55..901I, ZHUKOVSKA08, ASANO13, ROWLANDS14_2014MNRAS.441.1040R, ZHUKOVSKA14, DEVIS17_2017MNRAS.471.1743D, HOU19, AOYAMA20_2020MNRAS.491.3844A}.

In addition to dust growth, models show that star formation history \citep{ZHUKOVSKA14} and the change in dust size distributions \citep[e.g.\ coagulation and shattering;][]{HIRASHITA11,HIRASHITA+AOYAMA19,Relano20} also affect D/M. Simulations also suggest that the resolved D/M is correlated with a galaxy's gas fraction (\fgas, the fraction of gas mass to the total gas and stellar mass) and stellar mass distribution \citep{HOU19,LI19}. Thus, observing D/M across a range of environments can provide important constraints for dust evolution modeling.

One direct way to constrain D/M is to observe elemental depletions in the ISM \citep[i.e.\ the fraction of a given element in dust grains rather than in gas phase;][]{JENKINS87, JENKINS89}. Observations in \citet{JENKINS09} show that depletion increases with the ISM gas density along sightlines within the local part of the MW (distance < 10\,kpc). This also implies that D/M varies with ISM environment even when metallicity stays approximately the same. \citet{JENKINS17} and \citet{ROMAN-DUVAL19_2019ApJ...871..151R} also found a varying D/M in the Magellanic Clouds (MCs), where the metallicity is assumed to be approximately constant within each galaxy. These studies also showed that the depletion of dust forming elements, e.g.\ silicon and iron, increases with ISM gas surface density.

However, there are several limitations to the depletion observations. Most of them are due to the necessity of obtaining high resolution UV spectroscopy with high signal-to-noise ratio (S/N). These limitations include: (a) depletions are observable mainly in sightlines with relatively low dust extinction and moderate column densities; (b) depletions are only observable in galaxies where individual stars can be resolved or background quasars can be used; (c) some key constituents of dust grains, like carbon, are not observable outside the MW due to lack of current telescope facilities at the necessary wavelengths \citep{JENKINS17,ROMAN-DUVAL19_2019BAAS...51c.458R,ROMAN-DUVAL19_2019ApJ...871..151R}.

The other common method to determine D/M is to observe dust mass, gas mass, and metallicity separately, and then combine those observations. This method suffers from the combined systematic uncertainties in our understandings from various aspects of ISM physics, but it is still the best strategy we have except direct depletion measurements. The dust mass is usually derived from far-infrared (FIR) dust emission or near-infrared dust extinction \citep[][and references therein]{HILDEBRAND83_MBB,ISSA90, LISENFELD98,DRAINE07,COMPIEGNE11,DALCANTON15,JONES17_THEMIS,GALLIANO18}, while the gas surface density is derived from gas emission lines like the \HI 21\,cm \citep[e.g.][]{WALTER08} and CO rotational lines \citep[e.g.][]{LEROY09}. Two representative galaxy-integrated surveys using this strategy are \citet{REMY-RUYER14} and \citet{DEVIS19}. \citet{REMY-RUYER14} surveyed 126 galaxies and found that D/M increases with metallicity in galaxies with $\metal < 8.1$, and stays roughly constant in high-metallicity ones. On the other hand, the other survey across $\sim 500$ galaxies by \citet{DEVIS19} showed that D/M increases with metallicity across the entire observed metallicity range. \citet{DEVIS19} also showed that D/M correlates with other galaxy properties, e.g.\ stellar mass, specific star formation rate and \fgas. The exact dependence of D/M on galaxy properties remains controversial, which is at least partially a consequence that most of these quantities are mutually correlated.

Since most physical processes that affect D/M are associated with local ISM environments, spatially resolved D/M studies are necessary for constraining the dust models \citep{ZHUKOVSKA14,HU19} in addition to measuring galaxy-integrated D/M. There are several resolved studies targeting single or a few galaxies showing a varying D/M.  \citet{ROMAN-DUVAL14} and \citet{ROMAN-DUVAL17} found the dust-to-gas ratio (the ratio of dust surface density to total gas surface density, D/G) to increase with gas surface density at fixed metallicity in the MCs. \citet{CHIANG18} and \citet{Vilchez19} found the D/G to increase non-linearly with metallicity within the nearby spiral galaxy M101. On the other hand, \citet{DRAINE14} found a constant D/M in the disk of M31. One problem that emerges in comparing across these studies is the lack of uniformity. Different studies adopted different dust modelling, dust opacity, \CO-to-\HTWO conversion factor (\aco), and metallicity calibrations. All these factors together make it hard to compare previous D/M studies on an equal footing.

In addition to uniformity, these factors are also notorious for the level of disagreement among various methodologies. Several studies pointed out that dust opacity may vary. \citet{GORDON14} and \cite{CHIANG18} showed that the empirical opacity depends on the dust model under the same method of calibration. \citet{DALCANTON15} and \citet{PLANCK16} found that the dust mass estimated by the \citet{DRAINE07} dust model is $\sim 2$ times larger than the dust mass measured by extinction observations, suggesting a possible offset in dust opacity. \citet{Fanciullo15_2015A&A...580A.136F} estimated that dust opacity has a $\sim 20\%$ variation in the typical MW diffuse ISM. \citet{CLARK16} and \citet{CLARK19_2019MNRAS.489.5256C} showed that if D/M is fixed, dust opacity is inversely correlated with local ISM gas density, spanning a factor $\sim 8$. 

The \CO-to-\HTWO conversion factor, \aco\footnote{For the \CO-to-\HTWO column density conversion factor ($X_\CO$), a conversion that $X_\CO = 2\times 10^{20}\,\XcoUnit$ being equivalent to $\aco = 4.35$ \acoUnit\ is used throughout the paper. The mass of helium and heavy elements are included in the \aco factor.\label{lab: aCO}}, is known to vary with ISM environment, especially in low-metallicity regions, where the amount of \CO-dark \HTWO increases as the shielding from dust becomes weaker \citep{ISRAEL97,WOLFIRE10,LEROY11,GloverMacLow11,BOLATTO13,SANDSTROM13,HUNT15,Schruba17}. Several studies also find that \aco tends to be lower (2 to 10 times smaller than the disk-average value) in the centers of galaxies, possibly due to a stronger \CO emission in environments with higher temperature and gas turbulence \citep[][]{SANDSTROM13,Cormier18,Israel20}. Another problem regarding \aco selection for the purposes of measuring D/M is that many methods of measuring \aco have built-in assumptions of a fixed D/M or fixed D/G, which would not be self-consistent in studies of D/M variation. For more discussion regarding \aco, we refer our readers to the \citet{BOLATTO13} review and references therein. 

To determine metallicity accurately, the electron temperature (\Te) of the observed \HII region is required. \Te can be derived from temperature-sensitive auroral lines \citep[so called ``direct'' measurements, e.g.][]{BERG15}. However, the auroral lines are rarely used because their intensity is weak and thus hard to observe. The widely used ``strong line'' measurements make assumptions about \Te, and therefore have large systematic uncertainties between different calibrations \citep{KEWLEY_ELLISON08}.

In this work, we measure the spatially resolved D/M-environment relations in five nearby galaxies: IC342, M31, M33, M101, and NGC628. This selection is based on their distance and data availability (details in Sect.~\ref{sec: data}). By studying the resolved relation between D/M and local physical quantities across multiple galaxies, we can better constrain our understanding of the dust life cycle. We attempt to overcome the uniformity issues associated with previous studies by using the same calibrations of dust and metals. Moreover, we propose an approach to constrain the D/M and \aco simultaneously.

This paper is presented as follows. We describe our data and dust emission modelling in Sect.~\ref{sec: data}. In Sect.~\ref{sec: results}, we examine the D/M yielded from existing \aco prescriptions, and present a novel approach to constrain D/M and \aco simultaneously. We discuss the implications and interpretations of our D/M in Sect.~\ref{sec: discussions}. Finally, we present our conclusions in Sect.~\ref{sec: summary}.

\section{Sample and Data}\label{sec: data}

We study the D/M-environment relations in five nearby galaxies, IC342, M31, M33, M101, and NGC628. Their properties are tabulated in Table~\ref{tab: data}. We select these galaxies by the following criteria: (a) They have the photometry data of all five bands ranging $\lambda=100{-}500\,\micron$ observed by \textit{Herschel} PACS and SPIRE \citep{GRIFFIN10,PILBRATT10,POGLITSCH10} enabling uniform dust modeling. (b) They have both \HI and \CO maps available. (c) They have metallicity gradients derived from auroral line measurement in \HII regions. (d) Their distances are within 10\,\Mpc, which corresponds to a physical resolution better than 2\,\kpc at the coarsest resolution map (SPIRE 500\,\micron). Note that an exception is made for IC342 in the metallicity criteria because it has strong line metallicity measurements. We include it because it fits all other criteria. In addition, it spans the high SFR surface density (\Sigmasfr), high molecular gas surface density (\Sigmamol), and high gas volume density environments which are not covered by the other galaxies.

\input{tab_alldata}

We convolve maps from all selected galaxies to a uniform physical resolution using the \texttt{astropy.convolution} package \citep{ASTROPY13,Astropy18} and kernels from \citet{ANIANO11}. The common resolution for all multi-wavelength maps is defined as a Gaussian point spread function (PSF) with FWHM = 1.94\,kpc, which is equivalent to an angular FWHM = 41$\arcsec$ for our most distant galaxy, NCG628 \citep[here 41$\arcsec$ is the ``moderate'' Gaussian convolution for our coarsest resolution data, SPIRE 500;][]{ANIANO11}. After convolution, each map is then reprojected to a grid so that there are 2.5 pixels across the FWHM (i.e., we oversample at roughly the Nyquist sampling rate) using the \texttt{astropy} affiliated package \texttt{reproject}.

The IR and UV observations are blended with the cosmic background emission. To remove the background emission in the \textit{Herschel} maps, we follow the steps in \citet{CHIANG18}, which involves a tilted-plane fitting with iterative outlier rejection. For the WISE and GALEX maps, we use the data products from ``$z=0$ Multi-wavelength Galaxy Synthesis'' \citep[$z$0MGS,][]{LEROY19}, which have already been through background removal process.

To estimate the uncertainties of the observed quantities, we adopt the sensitivities or root-mean-square errors (rms) from the corresponding reference, multiplied by a factor of $\sqrt{N_f/N_i}$, where $N_f$ and $N_i$ are the numbers of resolution elements after and before convolution, respectively. Whenever there is only rms per channel available in the reference (e.g., \HI data in M101), we assume an average gas velocity dispersion $\sigma_{\rm z,\,gas}=11\,\rm km\,s^{-1}$ \citep{LEROY08} to calculate the integrated rms, that is:
\begin{equation}
    {\rm rms} = ({\rm rms~per~channel}) \times 2\sqrt{2\ln2}\ \sigma_{\rm z,\,gas}~,
\end{equation}
where the $2\sqrt{2\ln2}$ factor converts $\sigma_{\rm z,\,gas}$ to full width at half maximum.

We expect most quantities in this work to vary with galactocentric radius. The region above 3$\sigma$ detection is up to $\sim 0.8\R{25}$. For the galaxy with largest inclination, M31, the pixels near the minor axis and the center of the galaxy are severely blended with pixels in other radial regions after convolution. Thus, we blank M31 data in the $\pm 45^\circ$ region around the minor axis. The central $0.4\R{25}$ region of M31 is also blanked due to lack of metallicity data. The blanked region is shown in Figure~\ref{fig: dust maps}. All the surface density ($\Sigma$) terms presented in this work are corrected by a factor of $\cos(i)$ to account for inclination. This term will not be shown in the following equations.

\subsection{Dust mass}\label{sec: data-dust}
\subsubsection{\textit{Herschel} FIR data}\label{sec: data-dust-herschel}
We use the $\lambda=100{-}500\,\micron$ FIR images observed by the \textit{Herschel} PACS and SPIRE \citep{GRIFFIN10,PILBRATT10,POGLITSCH10} to derive dust properties. We use the $z$0MGS data products \citep[][J.~Chastenet et al.\ in preparation]{LEROY19}. The original observations were made by: 
IC342 \citep{KENNICUTT11(KINGFISH)2011PASP..123.1347K}, 
M31 \citep{FRITZ12(HELGA)2012A&A...546A..34F, GROVES12, DRAINE14}, 
M33 \citep{KRAMER10(HERMES)2010A&A...518L..67K, BOQUIEN11,XILOURIS12}, 
M101 \citep{KENNICUTT11(KINGFISH)2011PASP..123.1347K}, and 
NGC628 \citep{KENNICUTT11(KINGFISH)2011PASP..123.1347K}.

The native FWHMs are approximately $7\farcs0$, $11\farcs2$, $18\farcs2$, $24\farcs9$, and $36\farcs1$ for the 100, 160, 250, 350, and 500\,\micron band images, respectively. We do not include the $70\,\micron$ flux because the stochastic heating from small dust grains makes non-negligible contribution in that spectral range \citep[e.g.][]{DRAINE07}, which is not accounted for by the dust emission model we employ in this study.

\subsubsection{Fitting Dust Emission SED}\label{sec: data-dust-fitting}
We adopt a modified blackbody model (MBB) \citep{Schwartz1982,HILDEBRAND83_MBB} with a broken power-law emissivity to fit the dust emission spectral energy distribution (SED, represented by $I_\nu$) with the 100 to 500\,\micron \textit{Herschel} data. The free parameters in this model are dust surface density (\Sigmad), dust temperature (${\rm T}_d$) and the long-wavelength power-law index for emissivity ($\beta_2$). This model selection is based on the model comparison in our previous work. In \citet{CHIANG18}, we found the broken power-law emissivity MBB to yield a \Sigmad that is reasonably below the upper limit derived from metallicity, a ${\rm T}_d$ gradient matching the dust-heating environment, and one of the best $\chi^2$ value distributions among the five variants of MBB. In \citet{CHIANG18}, we have shown that the \Sigmad derived with a MBB with a broken power-law emissivity is within 0.1\,\dex of the \Sigmad derived with the commonly-used MBB with a constant power-law emissivity ($\beta$ fixed at 2.0).

The MBB model takes the form:
\begin{equation}\label{eq: MBB}
I_\nu(\lambda)\,[\SEDUnit] = \kappa(\lambda)\, \Sigma_d\, B_\nu(\lambda,T_d)~,
\end{equation}
where $\kappa(\lambda)$ is the wavelength-dependent emissivity, and $B_\nu(\lambda,T_d)$ is the blackbody SED at dust temperature $T_d$. We adopt a broken power-law emissivity \citep[][also see \citealt{REACH95}]{GORDON14,CHIANG18} described~by:
\begin{equation}\label{eq: BE}
\kappa(\lambda)=\left\{\begin{array}{ll}
    \kappa_{160}\left(\frac{\lambda_0}{\lambda_{\,}}\right)^{\beta} & \text{for}\quad \lambda < \lambda_b \\
    \kappa_{160}\left(\frac{\lambda_0}{\lambda_b}\right)^{\beta}\left(\frac{\lambda_b}{\lambda_{\,}}\right)^{\beta_2} & \text{for}\quad \lambda \geq \lambda_b
    \end{array}\right.,
\end{equation}
where $\lambda_b$ is the break wavelength fixed at 300\,\micron and $\beta$ is the short-wavelength power-law index fixed at 2.0; the long-wavelength power-law index ($\beta_2$) is left as a free parameter in the fitting\footnote{In this work, the fitted $\beta_2$ spans the 1-$\sigma$ range of $2.09^{+0.16}_{-0.22}$, $1.81^{+0.12}_{-0.29}$, $1.25^{+0.23}_{-0.26}$, $1.44^{+0.52}_{-0.42}$ and $1.84^{+0.28}_{-0.38}$ in IC342, M31, M33, M101 and NGC628, respectively. The overall 1-$\sigma$ range is $1.64^{+0.43}_{-0.50}$.}; $\lambda_0=160\,\micron$ is the reference wavelength for $\kappa(\lambda)$.

The reference emissivity $\kappa_{160}$ is calibrated with the depletion measurements and FIR SED in the MW cirrus \citep{JENKINS09, GORDON14}. The calibrated value for our model is $\kappa_{160}=20.73\pm 0.97\,\kappaUnit$ \citep{CHIANG18}. This calibration method is known to produce \Sigmad values not exceeding the upper bound given by the local available metals \citep{GORDON14,CHIANG18}.

We fit the dust SED in all pixels with $\text{S/N}>1$ in all five \textit{Herschel} bands following the grid-based fitting method presented in \citet{GORDON14} and \citet{CHIANG18}. We build a multidimensional grid with each grid point representing a combination of possible model parameters. At each pixel of the maps, we calculate the likelihood that a given model fits the observations, and repeat at all grid points. Finally, we compute the expectation values of the model parameters. The likelihood is calculated with a covariance matrix consisting of both variance of each band and the band-to-band covariance. This method allows us to directly account for the band-to-band correlation due to noise from astronomical sources, e.g.\ background galaxies and MW cirrus, which dominate the far-IR background noise. For more details, we refer to sect.~3.2 of \citet{CHIANG18} or sect.~4 of \citet{GORDON14}.

\begin{figure}[!htb]
\centerline{\includegraphics[width=1.0\columnwidth]{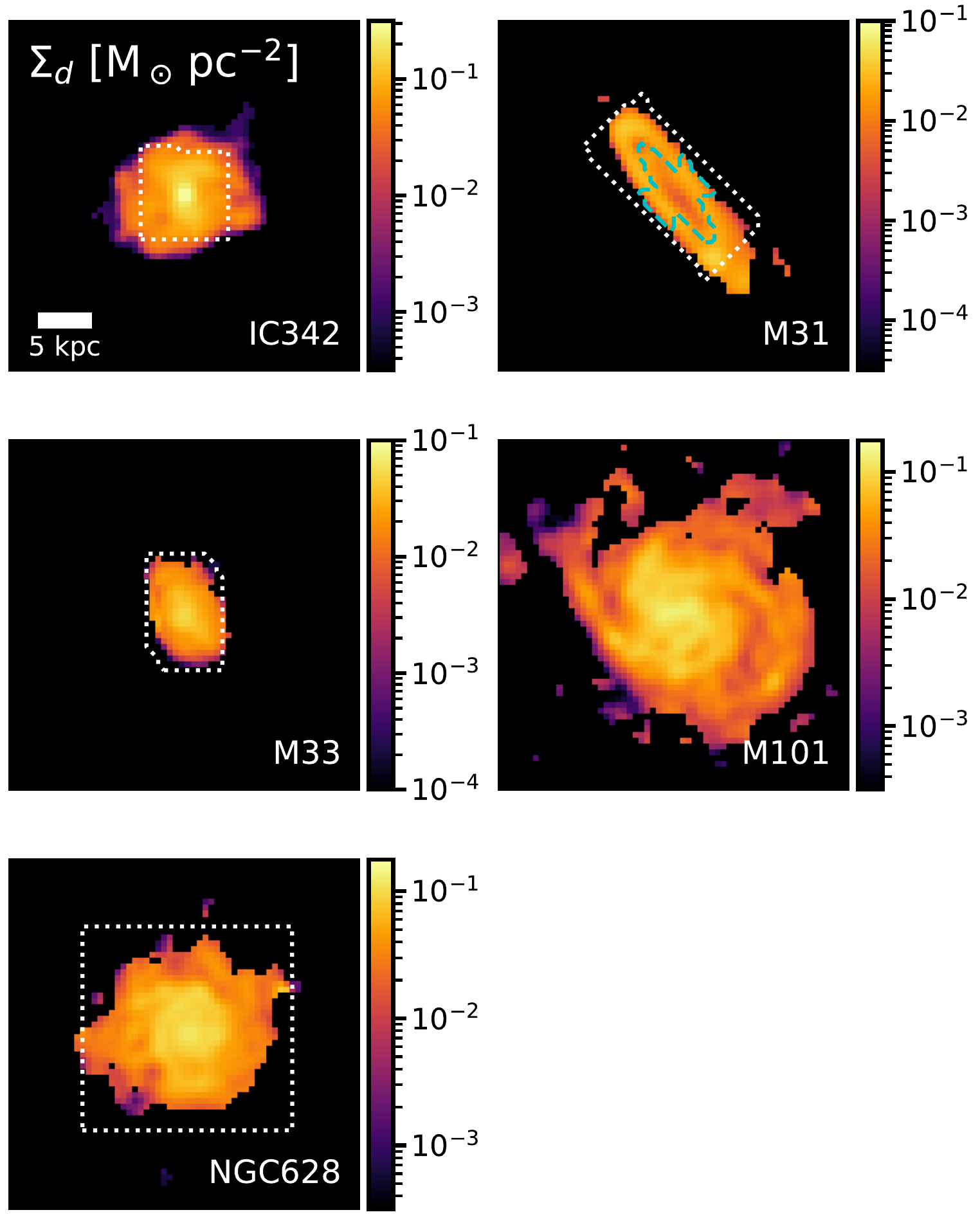}}
\caption{The fitted dust surface densities (\Sigmad [$\SigmaMassUnit$]) at the common resolution (a Gaussian PSF with FWHM=1.94\,\kpc). The levels of \Sigmad are indicated by the colorbars attached to each panel. Note that the scales in the panels are not identical. The white dotted line marks the boundary where we have detections in all observations, which we constrain our analysis to. This boundary is defined by \CO in most cases. The boundary of M101 is outside the plotting range. The cyan dashed line marks the region removed in M31 due to inclination and lacking metallicity data (see Sect.~\ref{sec: data}). The scale bar at the top left shows the 5 kpc length for all panels. The dust SED fitting is only performed in regions where $\text{S/N}>1$ in all five \textit{Herschel} bands. The region with $\text{S/N}\leq1$ in any of the five \textit{Herschel} bands appears in black. The $>1\,\dex$ range in \Sigmad within each galaxy indicates that we resolve the exponential disks at this resolution.\label{fig: dust maps}}
\end{figure}

The fitting is done at the common resolution. Figure~\ref{fig: dust maps} shows the resulting dust maps. Although the angular resolution has been degraded substantially for some galaxies, the range of $\Sigma_d$ at the common resolution is still more than one order of magnitude in each galaxy. This indicates the $\sim 2\,\kpc$ resolution resolves the exponential disks of our selected galaxies.

\subsubsection{Fitting errors}\label{sec: data-dust-err}
For each model parameter ($X=\logt\Sigmad$, $T_d$, or $\beta_2$), we use the marginalized likelihood-weighted 16-/84-percentile ($X_{16}$ and $X_{84}$, respectively) at each pixel to represent the $1{\text -}\sigma$ distribution. We then quote the maximum difference between the expectation value ($X_{\rm exp}$) and the $1{\text -}\sigma$ distribution as the fitting error $\epsilon_X$, that is:
\begin{equation}\label{eq: fit err}
    \epsilon_X = \max\left((X_{84} - X_{\rm exp}),\ (X_{\rm exp} - X_{16})\right)~.
\end{equation}
This is the same method as in \citet{CHIANG18}.

\subsection{Gas masses}\label{sec: data-gas}
We calculate the total gas surface density (\Sigmagas) as:
\begin{equation}\label{eq: total gas}
    \Sigmagas = \Sigmaatom + \Sigmamol~,
\end{equation}
where \Sigmaatom is the atomic gas surface density and \Sigmamol is the molecular gas surface density.

\subsubsection{Atomic gas mass}\label{sec: data-gas-hi}
We use new and archival \HI 21\,cm line emission ($I_\HI$) data to trace \Sigmaatom. The data sources are: IC342 (P.I.\ K.~M.\ Sandstrom; I.\ Chiang et al.\ in preparation)\footnote{Observed with the \textit{Karl G. Jansky} Very Large Array (VLA).}, M31 \citep{BRAUN09}, M33 \citep{KOCH18}, M101 \citep{WALTER08}, and NGC628 \citep{WALTER08}. The resolution of the \HI data is always high enough that it never limits our analysis. For M31 and M33, the two galaxies with largest angular scales, a short-spacing correction with GBT data has been included in the original works \citep{BRAUN09, KOCH18}. The short-spacing correction is not applied in IC342, M101 and NGC628.

Among these three galaxies, IC342 is most likely to have its $I_\HI$ underestimated with interferometric data only due to its sky coverage: the \HI 21-cm signal in IC342 spans a diameter $\sim 45\arcmin$, whereas the largest angular scale covered by the VLA D-configuration is $\sim 16\arcmin$ in the L band. The total atomic mass in our IC342 map is $\rm M_{atom} = 7.9 \times 10^9~M_\odot$, which is close to the $\rm M_{atom} = 8.4 \times 10^9~M_\odot$ in \citet[][distance corrected]{Crosthwaite00}. Single-dish measurement in literature \citep[][distance corrected]{Rots79} showed a $\rm M_{atom} = 18.7 \times 10^9~M_\odot$. However, this value is expected to be overestimated because the low spatial and velocity resolutions in \citet{Rots79} data are not enough to distinguish and remove the MW foreground completely.

We also compare our result with the recent single-dish data \citep[EBHIS,][]{Kerp11_EBHIS,Winkel16_EBHIS}. We choose a spectral range that is free from the MW foreground, $\sim 43{-}128$~km/s (the \HI 21-cm signal in IC342 spans $\Delta v\sim 210$~km/s). We find the total flux from EBHIS data is $\sim 1.6$ times larger than our VLA measurement in this range, which indicates that a short-spacing correction is desired but confused by MW foreground emission. Since this 1.6 factor is an average value instead of an offset that can be directly applied to all pixels, we do not include it in our analysis. This factor does not affect our main conclusions due to the low atomic gas content in our region of interest, which is at the center of IC342.

We calculate \Sigmaatom from $I_\HI$ via the following equation, assuming the opacity is negligible \citep[e.g.][]{WALTER08}:
\begin{multline}\label{eq: hi}
    \Sigmaatom\,[\SigmaMassUnit] = \\ 1.36\times(8.86\times10^3)\times \left(\frac{I_\HI\,[\IHIUnit]}{\Bmaj\,[\arcsec]\times \Bmin\,[\arcsec]}\right)~,
\end{multline}
where \Bmaj and \Bmin are the full-width half-maximum (FWHM) of the major and minor axes of the synthesized beam, respectively. The $1.36$ factor accounts for the mass of helium and heavy elements.

\subsubsection{Molecular Gas Mass}\label{sec: data-gas-mol}
We use \CO rotational line emission ($I_\CO$) to trace \Sigmamol. The data sources are: IC342 (A.~Schruba et al.\ in preparation)\footnote{Observed with the NOrthern Extended Millimeter Array (NOEMA) and the IRAM \mbox{30-m} telescope.}, M31 \citep{NIETEN06}, M33 \citep{GRATIER10,DRUARD14}, M101 \citep{LEROY09}, and NGC628 \citep{LEROY09}. The \CO resolution is always high enough that it never limits our analysis.

Throughout the paper, \aco is quoted for the \COonezero rotational line at 115\,\GHz and includes a factor to account for helium. However, we use the 230\,\GHz \COtwoone data in M33, M101, and NGC628. In those cases, we quote a \mbox{(2-1)}$/$\mbox{(1-0)} brightness temperature ratio (\R{21}) to convert the integrated intensity, that is:
\begin{equation}
    I_\COonezero\,[\ICOUnit] = \frac{I_\COtwoone\,[\ICOUnit]}{\R{21}}~.
\end{equation}
We use $\R{21}=0.8$ in M33 \citep{GRATIER10,DRUARD14}, and $\R{21}=0.7$ in M101 and NGC628 \citep{LEROY13}. We do not include uncertainties resulting from variations of \R{21} in the analysis. The uncertainty in D/M due to the choice of \R{21} is $\leq 0.05\,\dex$ in the three galaxies using \CO $\rm J=2\to1$ data (M33, M101 and NGC628). In M33 and NGC628, the measured variations of \R{21} are reasonably small \citep{SANDSTROM13,DRUARD14}. While in M101, \R{21} could increase by $\sim0.3$\,\dex in the central $\sim 0.05$~\R{25} \citep[$\sim 1.2$~kpc,][]{SANDSTROM13}, which indicates that we might underestimate \Sigmamol in that small region. The \R{21} values adopted in this study are consistent with the PHANGS measurements ($\R{21}=0.64\pm0.09$, J.~den Brok et al. 2020, A\&A submitted) considering the systematic uncertainties due to calibration \citep[e.g., 15\% for \CO $\rm J=2\to1$ data in][]{DRUARD14}.

We can translate $I_\CO$ to \Sigmamol via a \CO-to-\HTWO conversion factor (\aco):
\begin{multline}\label{eq: h2}
    \Sigmamol\,[\SigmaMassUnit] = \\
    \aco\,[\acoUnit]\times I_\COonezero\,[\ICOUnit]~.
\end{multline}

\input{tab_aco_prescriptions}

Since D/M is sensitive to the choice of \aco, we calculate our results with four \aco prescriptions in this study (see Table~\ref{tab: aco_prescription}). The conventional MW \aco \citep[\acoMW;][]{SOLOMON87_1987ApJ...319..730S,STRONG96_1996AA...308L..21S,ABDO10_2010ApJ...710..133A} is one of the most widely used choices for \aco. It has a fixed value at 4.35\,\acoUnit and no dependence on the environments (see footnote \ref{lab: aCO} for the conversion between \aco and $X_\CO$). The \citet[][their table~7, the ``all galaxies'' formula with HERACLES sample]{SCHRUBA12} prescription (\acoSchruba) models \aco as a simple power law with metallicity, which is a common strategy in modelling \aco \citep[e.g.][]{ISRAEL97,FELDMANN12,HUNT15,ACCURSO17}. \acoSchruba has the largest normalization factor among the prescriptions here, thus it results in the overall highest \Sigmagas, or the smallest D/M. Another power-law prescription we include here is the \citet[][Sect.~5.1]{HUNT15} prescription (\acoHunt), which is a power law with metallicity in regions below solar metallicity ($\rm Z_\odot$) and a constant at \acoMW above $\rm Z_\odot$. This cut off is due to smaller amount of \CO-dark \HTWO at high metallicity. The \citet[][Eq.~31]{BOLATTO13} prescription (\acoBolatto) has an exponential dependence on metallicity and a power law dependence on total surface density (\Sigmatot = \Sigmagas + \Sigmastar) in the regions with highest surface densities. Note that we assume $\Sigma_{\rm GMC}=100\,\SigmaMassUnit$ in the \acoBolatto case.

\subsection{Metallicity}\label{sec: data-metal}
\input{tab_metaldata}

We use the oxygen abundance, \metal, to trace metallicity in this work. We adopt $\metal_\odot=8.69$ \citep{ASPLUND09}. We adopt measurements from multiple sources (Table~\ref{tab: metal}). We use gradients of \metal derived from auroral line measurements in \HII regions in all galaxies except IC342. For IC342, we use the S-calibration for strong lines from \citet[][hereafter PG16S]{PilyuginGrebel16}, which is a calibration showing good agreement with direct metallicity measurements \citep{CROXALL16,Kreckel19}. Within the region of interest in this work, the \metal ranges from 8.2 to 8.8. Note that in M31, the metallicity gradient is derived with data outside 0.4\,\R{25} only \citep{ZURITA12}, thus we blank all M31 data within 0.4\,\R{25} in the D/M analysis.

In the calculation of D/M, we need to convert the \metal to metallicity ($Z=\Sigmametal/\Sigmagas$, note that \Sigmagas includes the mass of heavy elements in our notation) because a complete measurement of abundance of all elements is unavailable. We use a fixed oxygen-to-metals ratio, $\rm M_O/M_{metal}=0.51$, calculated from the solar neighborhood chemical composition \citep{LODDERS03}. The complete conversion is:
\begin{equation}\label{eq: O/H-to-metal}
    Z = \frac{1}{\rm M_O/M_{metal}}\frac{\rm m_O}{\rm 1.36 m_H} 10^{\left(\metal\right) - 12}~,
\end{equation}
where $\rm m_O$ and $\rm m_H$ are the atomic masses for oxygen and hydrogen, respectively; the 1.36 factor converts hydrogen mass to total gas, which is consistent with the conversion in Sect.~\ref{sec: data-gas}. We do not include a correction of [O/H] due to depletion of oxygen in \HII regions, which is estimated to be $\lesssim 0.1\,\dex$ \citep{Esteban98,Peimbert10}.

Although we do our best to quote the most reliable metallicity, we would like to remind the readers of two remaining caveats in our methodology: (a) A~fixed oxygen-to-metals ratio across all ISM environments might not be true considering the variation of chemical composition in the ISM \citep[e.g.\ the variation of \logt(N/O) in][]{CROXALL16}. Currently, there is no good observational method to characterize this ratio for all environments. Simulation results suggest that it is reasonable to treat it as a constant at this point \citep[e.g.][]{MA16}. (b) We use metallicity gradients instead of a complete metallicity map, which might cause an artificial correlation between D/M and galactocentric radius. In massive spiral galaxies, the variation of metallicity is dominated by the radial gradient, and the azimuthal scatter is considered second order. \citet{CROXALL16} and \citet{BERG15} measure representative azimuthal scatter of $\sim0.1\,\dex$ in M101 and NGC628, which is small compared to the radial gradient but non-negligible. \citet{Kreckel19} found that the typical scatter of \metal at a given radius in the PHANGS-MUSE samples is small, which is around 0.03 to 0.05\,\dex. There are ongoing efforts in fitting a complete \metal map from sightlines of \HII regions (T.~Williams et al.\ in preparation). Their preliminary results also show that the radial gradient dominates the variation in \metal.

\subsection{Star Formation Rate and Stellar Mass}\label{sec: data-SFR}
We use the GALEX \citep{MARTIN05} and WISE \citep{WRIGHT10} maps to trace star formation rate surface density (\Sigmasfr) and stellar mass surface density (\Sigmastar). For both GALEX and WISE maps, we use the $z$0MGS data products \citep{LEROY19} with a resolution of 15\arcsec. The correction for the MW extinction in the GALEX maps have been included in the $z$0MGS data products.

The continuum at the GALEX FUV band ($\sim 154\,\nm$) is dominated by the light from relatively young ($\lesssim 100\,\Myr$) stars, so we can estimate \Sigmasfr from the FUV flux ($I_{\rm FUV}$). Since interstellar dust absorbs the starlight and re-emits it in the IR \citep{CALZETTI07,KENNICUTT12}, we further improve the estimation by correcting the $I_{\rm FUV}$ with local dust extinction using WISE W4 ($\sim 22\,\micron$) flux ($I_{\rm W4}$). We adopt the hybrid SFR calibrations in table~7 of \citet{LEROY19}:
\begin{multline}
\Sigmasfr\,[\SigmasfrUnit] \approx \\
8.85 \times 10^{-2} I_{\rm FUV}\,[\SEDUnit] + 3.02 \times 10^{-3} I_{\rm W4}\,[\SEDUnit]~.
\end{multline}
Note that although we adopt GALEX FUV maps that have been corrected for the MW extinction, the IC342 \Sigmasfr derived from GALEX FUV could be uncertain due to its high MW extinction. However, the impact is small since the \Sigmasfr is dominated by the WISE W4 term.

We use the WISE W1 ($\sim3.4\,\micron$) maps to trace \Sigmastar. We adopt a stellar-to-W1 mass-to-light ratio, $\Upsilon^{3.4}_\star$, from $z$0MGS\footnote{$\Upsilon^{3.4}_\star=0.21$, 0.5, 0.29, 0.28 and 0.31 for IC342, M31, M33, M101, and NGC628, respectively.} \citep{LEROY19}. We then use this $\Upsilon^{3.4}_\star$ to calculate \Sigmastar from WISE W1 flux ($I_{\rm W1}$):

\begin{equation}
\Sigmastar\,[\SigmaMassUnit] \approx 3.3 \times 10^2 \left(\frac{\Upsilon^{3.4}_\star}{0.5}\right) I_{\rm W1}\,[\SEDUnit]~.
\end{equation}

\subsection{Ratios and Fractions}\label{sec: data-derived}
We use three derived ratios and fractions in the following analysis, which are calculated with the following formulas:
\begin{equation}
    {\rm D/M} \equiv \frac{\Sigmad}{\Sigmametal} = \frac{\Sigmad}{\Sigmagas\times Z}~,
\end{equation}
\begin{equation}
    \fhtwo \equiv \frac{\Sigmamol}{\Sigmagas}~,
\end{equation}
\begin{equation}
    \fgas \equiv \frac{\Sigmagas}{\Sigmagas + \Sigmastar}~.
\end{equation}
Note that whenever we calculate the galaxy-averaged or radial-binned values of these quantities, we calculate the ratio of averages, instead of the average of ratios.

\subsection{Dynamical Equilibrium Pressure}\label{sec: data-pressure}
We use the mid-plane dynamical equilibrium pressure (\PDE) to trace the volume density of gas in the ISM. We estimate \PDE with the same basic formulism in \citet{Elmegreen89}, \citet{LEROY08}, \citet{Gallagher18} and \citet{Sun20}:
\begin{equation}
    \PDE = \frac{\pi G}{2}\Sigmagas^2 + \Sigmagas\sqrt{2G\rho_\star}\sigma_{\rm gas,\,z}~.
\end{equation}
The first term represents the weight of the ISM due to the self-gravity of the ISM disk (\PDEself). The second term is the weight of the ISM due to stellar gravity (\PDEstar). $\sigma_{\rm gas,\,z}$ is the vertical gas velocity dispersion. We adopt a constant value of $\sigma_{\rm gas,\,z}=11\,\rm km\,s^{-1}$ from \citet{LEROY08}. $\rho_\star$ is the stellar mass volume density near the mid-plane. We estimate $\rho_\star$ with:
\begin{equation}
    \rho_\star = \frac{\Sigmastar}{4H_\star} =  \frac{\Sigmastar}{0.12\R{25}}~,
\end{equation}
where $H_\star$ is the stellar scale height. We estimate $H_\star$ with a fixed flattening ratio $\R{25}/H_\star=33.6$ \citep{LEROY08,Sun20}. The systematic uncertainty in \PDE resulting from the adopted $\sigma_{\rm gas,\,z}$ and the \R{25}-to-$H_\star$ conversion is $\sim 0.1{-}0.2$~\dex \citep{LEROY08,Sun20}. This is not included in the following Monte Carlo analysis since it is not a random error.

\section{D/M and the CO-to-H\texorpdfstring{$_2$}{2} Conversion Factor}\label{sec: results}
In this section, we propose a novel approach to constraining D/M and \aco simultaneously by examining the resolved environmental dependence of D/M on metallicity and ISM gas density. We expect that if all relevant quantities are accurately measured, we should observe D/M to increase or stay constant with both metallicity and ISM gas density. It has been demonstrated in several depletion-based D/M studies that D/M is positively correlated with both metallicity and gas volume density \citep{JENKINS09,JENKINS14,ROMAN-DUVAL19_2019ApJ...871..151R,PerouxHowk20}. From a theoretical perspective, it is also shown that if dust growth dominates over other dust input mechanisms, D/M would be positively correlated with both metallicity and ISM gas density; if the dust growth rate is lower due to either low dust or gas-phase metals abundance, D/M would stay roughly constant \citep{HIRASHITA99, INOUE03_2003PASJ...55..901I, ZHUKOVSKA08, ASANO13, ROWLANDS14_2014MNRAS.441.1040R, DEVIS17_2017MNRAS.471.1743D, HOU19, AOYAMA20_2020MNRAS.491.3844A}.

We take \metal and \PDE as tracers for metallicity and ISM gas density, respectively. We calculate the Pearson correlation coefficients of the radial dependence to quantify the D/M-\metal and D/M-\PDE correlations. In Sect. \ref{sec: results-D/M+aco}, we first calculate D/M with four existing \aco prescriptions, and examine the D/M-\metal and D/M-\PDE correlations. In Sect.~\ref{sec: results-constrain}, we attempt to constrain D/M and \aco simultaneously with the expected correlations. In Sect.~\ref{sec: results-summary}, we summarize the results in the above two sections. We show the profiles of all measurements calculated with \acoBolatto in Appendix~\ref{app: results-other}.

\subsection{Inspecting \texorpdfstring{\aco}{aCO} Prescriptions}\label{sec: results-D/M+aco}
\begin{figure*}[!ht]
\centerline{\includegraphics[width=\textwidth]{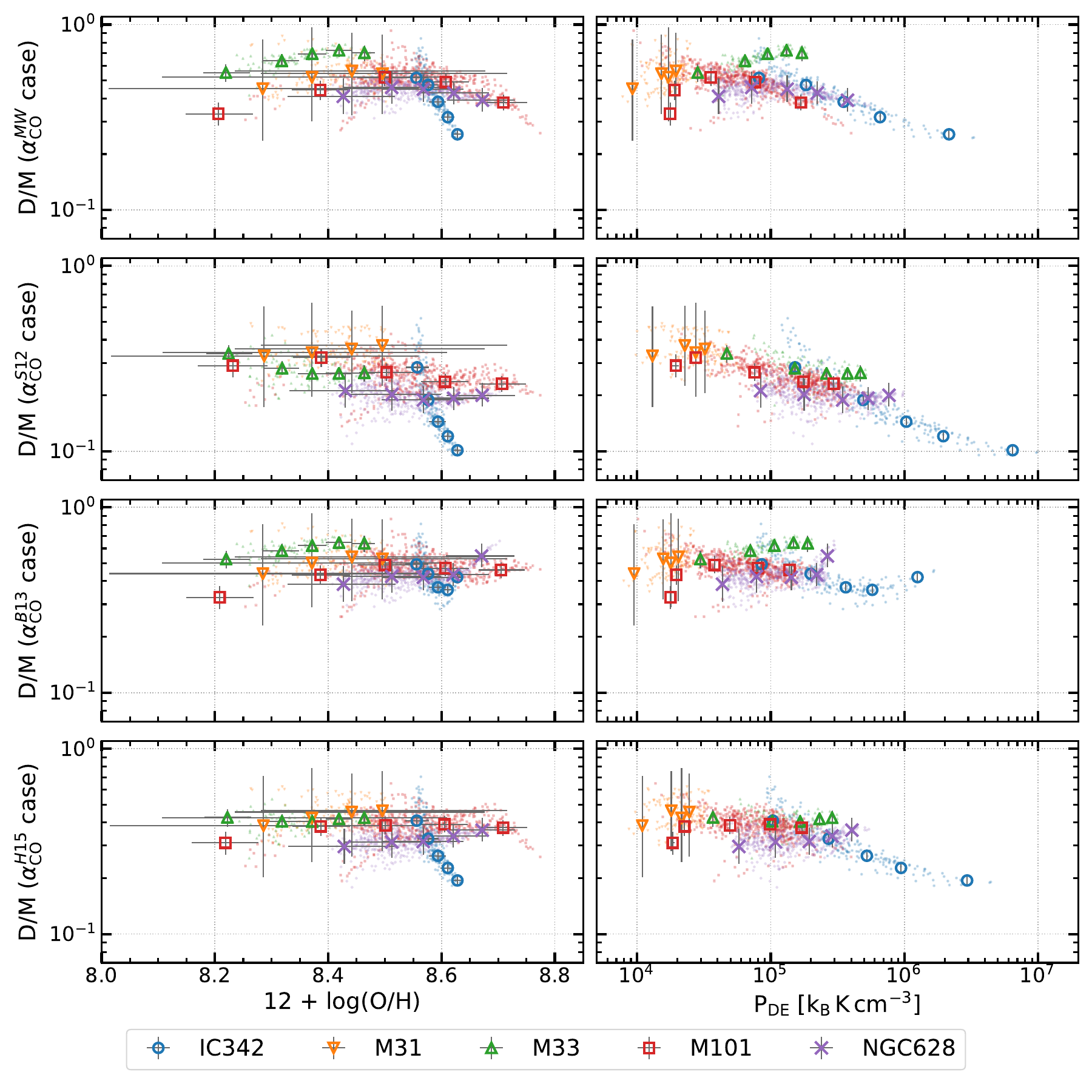}}
\caption{Measured D/M as calculated with four different \aco prescriptions. The large symbols are radial-binned values, while the small, faint symbols are the pixel-by-pixel values where detection is above $3\sigma$. The errorbar shows the 16-/84-percentile distribution from 1000 Monte Carlo tests assuming Gaussian error in measurements. The $x$-axes are \metal and \PDE, which are our tracers for metallicity and gas volume density, respectively.\label{fig: MetalDensity}}
\end{figure*}

\input{tab_metaldensity_corr_MonteCarloRadial}

We calculate \Sigmagas and D/M with four widely used \aco prescriptions (Sect.~\ref{sec: data-gas-mol}), and examine their resulting D/M-environment relations. In Figure~\ref{fig: MetalDensity}, we show D/M versus \metal and \PDE calculated from the four \aco prescriptions. The Pearson correlation coefficients of the radial trends within each galaxy are presented in Table~\ref{tab: metaldensity_corr}. The variances of the correlation coefficients are derived with the 16-/84-percentiles from 1000 Monte Carlo tests, assuming Gaussian uncertainties in \Sigmad, \Sigmastar, \Sigmaatom, $I_\CO$, and coefficients in the \metal gradients.

In Figure~\ref{fig: MetalDensity}, we notice that IC342 deviates from the other galaxies in the D/M-\metal trend except for \acoBolatto. We also notice that M31 has large uncertainties in D/M, mainly due to its uncertainties in the metallicity gradient, which makes its correlation coefficients in Table~\ref{tab: metaldensity_corr} less meaningful.

If we put IC342 and M31 aside for a moment, we find all \aco prescriptions except \acoSchruba to yield reasonable D/M-\metal and D/M-\PDE correlation coefficients. Meanwhile, the correlation coefficients are sensitive to the choice of \aco. \acoMW yields significant positive or insignificant D/M-\metal and D/M-\PDE correlations. \acoSchruba yields significant negative correlations in M33 and M101, and insignificant correlations in NGC628. \acoBolatto yields significant positive correlations. \acoHunt yields significant positive correlations in M101 and NGC628, and insignificant correlations in M33.

In IC342, we observe strong negative correlations with small variances with \acoMW, \acoSchruba and \acoHunt. \acoBolatto yields weaker and less significant negative correlations. Meanwhile, the D/M-\metal trend in IC342 stays within the range among the other galaxies with \acoBolatto in Figure~\ref{fig: MetalDensity}. One possible reason for the distinct behavior of IC342 is the starburst region in its center, which could affect dust SED fitting and \aco due to its temperature and gas velocity dispersion. Regarding the dust SED fitting in the center, we have a fairly good fit quality ($\chi^2\lesssim 1$) and a derived dust temperature ($T_d\sim 25\,\rm K$) that can be well described by an MBB within $\lambda=100-500\,\micron$, thus we trust our derived \Sigmad. Among the \aco prescriptions, \acoBolatto is the only one that considers the decrease of \aco due to gas temperature and gas dynamics. These effects are modelled by \Sigmatot\footnote{\citet{BOLATTO13} uses \Sigmatot to model the effects from gas temperature and gas dynamics for two reasons: (i) The temperature and velocity dispersion effects are more important in galaxy centers and Ultra Luminous Infrared Galaxies (ULIRGs). These regions can be captured by \Sigmatot with a lower bound in general. (ii) \Sigmatot is more easily measurable than the temperature and velocity dispersion.} in \citet{BOLATTO13}. This consideration likely results in the least negative D/M-\metal and D/M-\PDE correlation coefficients.

In summary, given that we expect D/M to increase or stay constant with both \metal and \PDE, \acoBolatto seems to give the most reasonable D/M among the four prescriptions across all environments. The D/M calculated with \acoBolatto has a mean value of 0.46, and a 1-$\sigma$ range spanning $0.40{-}0.58$. \acoMW and \acoHunt yield reasonable correlations in M33, M101 and NGC628, but result in strong negative correlations in IC342. Two effects likely contribute to this: (i) the distinct behavior of \aco due to the high \Sigmagas and \Sigmastar, which is not considered in \aco prescriptions parameterized by metallicity only, e.g., \acoSchruba and \acoHunt; and (ii) due to the high \fhtwo in IC342, the variation of \aco has a larger impact on D/M, \PDE, and their relevant correlations.

\subsection{Constraining \texorpdfstring{\aco}{aCO} with \texorpdfstring{D/M-\metal and D/M-\PDE
}{D/M-12+log(O/H) and D/M-PDE} Relations}\label{sec: results-constrain}
\begin{figure*}[!ht]
\centerline{\includegraphics[width=0.85\textwidth]{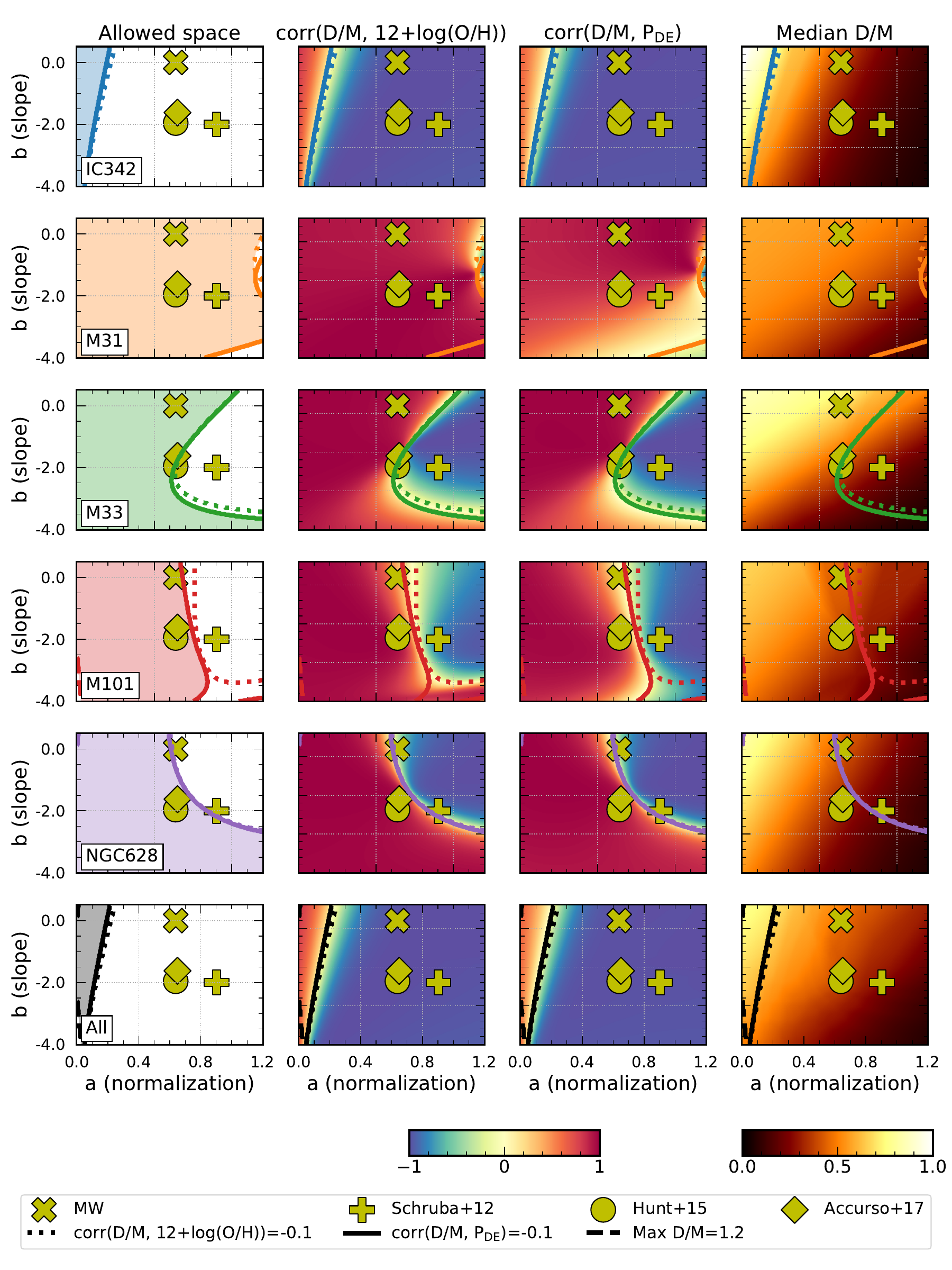}}
\caption{Power-law parameter space with the three constraints in each galaxy. In the ``allowed space'' column, the shaded area shows the region where all three constraints are satisfied. The x, plus, circle and diamond symbols mark the locations of \acoMW, \acoSchruba, \acoHunt and the \citet{ACCURSO17} \aco in the parameter space, respectively. The dotted, solid and dashed lines show the boundaries of the D/M-\metal correlation, D/M-\PDE correlation and the D/M upper limit constraints, respectively. In the central two columns, the colormap shows the Pearson correlation coefficients. In the ``Median D/M'' column, the colormap shows the pixel-by-pixel median D/M. To fit the 2-dimensional space, we assume $Z<Z_\odot$ for \acoHunt, and assume $\rm \Delta MS = 0$ for the \citet{ACCURSO17} prescription. All \aco prescriptions shown here locate outside the space satisfying all constraints, which implies that extra parameters are needed in \aco modeling to obtain a reasonable D/M across physical environments.\label{fig: aco_hist2d}}
\end{figure*}

We demonstrated that the resolved behavior of D/M is sensitive to the assumed conversion factor. Here, we propose a novel approach to constrain \aco by the expected D/M-metallicity and D/M-ISM gas density behaviors, which aims for solving D/M and \aco simultaneously. In the following, we present a first attempt at using this novel method to study the parameter space for the widely-used simple metallicity power-law prescriptions for \aco.

We model \aco as a simple power-law parameterized with metallicity, that is,
\begin{multline}
    \logt\frac{\aco}{1\,\acoUnit} = \\
    a + b \times \left(\metal - 8.69\right)~.
\end{multline}
We then constrain the parameter space $[a,\,b]$ to only include the non-negative D/M-\metal and D/M-\PDE correlations. We further constrain the parameter space with $\text{D/M}<1$ to ensure the  sanity of the resulting prescription. Practically, we relax the lower bound of the correlation coefficient to $\rho>-0.1$ to compensate for uncertainties in measurements. For the same reason, we relax the maximum D/M to $1.2$.

We explore the parameter space $0.0 \leq a \leq 1.25$, which is equivalent to a normalization of $0.25 \leq \aco/\acoMW \leq 4.0$ at solar metallicity. The range of $b$ is $-4 \leq b \leq 0.5$, which generously encompasses the slopes from previous extragalactic studies, which typically find $\aco\propto Z^{-1}$ to $Z^{-3}$ \citep{BOLATTO13}.

The constraints are visualized in the power-law parameter space in Figure~\ref{fig: aco_hist2d}. In the ``allowed space'' column, we see the maximum D/M constraints the minimum normalization so that the resulting \aco does not yield unphysical D/M in galaxy centers. The D/M-\metal and D/M-\PDE correlations constrain the maximum normalization at fixed slope. They limit the upper bound of \Sigmagas, thus the lower bound of D/M, in galaxy centers, so the \aco does not yield negative correlations. The boundary drawn by the two correlation constraints are similar to each other, while the constraint set by the D/M-\PDE correlation is usually more strict than the D/M-\metal correlation constraint.

Among the galaxies, IC342 has the narrowest allowed space, which primarily defines the parameter space allowed in all galaxies. The median D/M in this allowed space is high within IC342 and across all galaxies, implying that this space satisfies all constraints by minimizing \Sigmagas and creating a high, flat D/M. In M31, all constraints are easily satisfied within the parameter space we explore. M33, M101, and NGC628 have a large overlapping region in the allowed parameter space. The D/M upper limit constraint in M101 marks part of the boundary of the parameter space allowed in all galaxies.

We overlay four power-law or power-law-like \aco prescriptions, i.e.\ \acoMW, \acoSchruba, \acoHunt, and the \citet{ACCURSO17} prescription, on the parameter space. For \acoHunt, we only plot its low-metallicity solution. The complete \acoHunt formula would be the space between \acoMW and \acoHunt in Figure~\ref{fig: aco_hist2d}. We show \citet{ACCURSO17} because it is a widely-used power-law \aco. We did not include it in the previous analysis because it yields results similar with \acoSchruba. To fit the \citet{ACCURSO17} into the 2-dimensional space, we assume $\rm \Delta MS = 0$ in their eq.~25. In M33, M101, and NGC628, these prescriptions sit near the boundary of the correlation constraints. In IC342, these prescriptions are far from the allowed space.

The space that satisfies all constraints in all galaxies (bottom left panel in Figure~\ref{fig: aco_hist2d}) has a small normalization and a flat slope. The normalization spans $0.2 \lesssim \aco/\acoMW \lesssim 0.3$ at solar metallicity, and the slope spans $-1 \lesssim b \lesssim 0.5$. Although we do find a parameter space where all constraints are satisfied, it is almost solely defined by IC342. We do not proceed the D/M analysis with this parameter space as it yields a median D/M that seems to high compared to depletion observations.

The results from this test demonstrate that in galaxies except IC342, a simple power law \aco parameterized with metallicity can yield expected physics. Among the tested existing \aco prescriptions, the \acoMW, \acoHunt, and the \cite{ACCURSO17} satisfy most constraints in galaxies except IC342, while \acoSchruba seems to have a normalization that is too high (2\acoMW at $Z_\odot$). When we include IC342, the only space that satisfies all constraints yields D/M that seems too high, and the tested existing \aco prescriptions are far outside this allowed space. This suggests that one would probably need a more sophisticated functional form to properly model \aco across all environments (e.g., the starburst region in IC342). One example is the \citet{BOLATTO13} prescription, where the authors attempt to model the decrease of \aco in the high-\Sigmatot regions due to the combined effects of gas temperature and velocity dispersion.

\subsection{Section Summary}\label{sec: results-summary}
We showed that the D/M is sensitive to the choice of \aco. Among the prescriptions in Sect.~\ref{sec: results-D/M+aco}, \acoBolatto gives the most reasonable D/M. This is inferred from the D/M-\metal and D/M-\PDE correlations, especially in IC342. In Sect.~\ref{sec: results-constrain}, we use a new approach to constrain D/M and \aco simultaneously. However, in this first attempt, we show that the \aco satisfying all constraints yields D/M too high compared to depletion observations. Thus, we proceed with the \acoBolatto case for the following analysis.

The median and the 16-/84-percentile of our observed D/M calculated with \acoBolatto is $0.46^{+0.12}_{-0.06}$. This is consistent with the values adopted in \citet{CLARK16,CLARK19_2019MNRAS.489.5256C}, which are $0.5\pm0.1$ and $0.4\pm0.2$, respectively. The median D/M and the 16-/84-percentile in each galaxy are $0.41^{+0.11}_{-0.05}$, $0.50^{+0.11}_{-0.06}$, $0.60^{+0.05}_{-0.06}$, $0.48^{+0.06}_{-0.05}$ and $0.43^{+0.03}_{-0.04}$ for IC342, M31, M33, M101 and NGC628, respectively. Due to our limited understanding of the \CO-to-\HTWO conversion factor, we cannot conclusively determine the environmental dependence of D/M. We present the observed environmental dependence calculated with \acoBolatto in Appendix~\ref{app: results-other}.

\section{Discussion}\label{sec: discussions}
\subsection{Implications of the Observed D/M}\label{sec: discussions-implications}
In Sect.~\ref{sec: results}, we calculate D/M with several common \aco prescriptions and the parameter space of a power-law \aco parameterized by metallicity. Although we have not fully explored all possible descriptions of \aco, we proceed the analysis with the most reasonable prescription, \acoBolatto, at the moment. The \acoBolatto yields a fairly constant D/M over a wide range of physical environments, with a median $\rm D/M=0.46$. From dust evolution simulations \citep{DWEK98,ASANO13,AOYAMA20_2020MNRAS.491.3844A}, one possible explanation for a constant D/M is that dust growth dominates the increase of D/M, and the dust growth rate slows down as the available dust-forming metals in gas phase decreases. Thus, D/M would stay roughly constant when most dust-forming metals are already locked in dust grains.

This idea can be demonstrated with the toy model in \citet{ANIANO20}, which considers dust growth in the ISM, dust injection from AGB stars and supernovae, and dust destruction. It is assumed that the effective dust growth time scale ($\tau_a$) is much smaller than the dust injection time scale ($\tau_\star$), thus this model only applies to ISM environments where dust formation is dominated by dust growth in the ISM. With a quasi-steady state assumption, the model gives D/M as a function of metallicity:
\begin{equation}
    {\rm D/M} = 0.5\frac{f_m}{Z'}\left\{\left(Z' - \frac{\tau_a}{\tau_d}\right) + \left[\left(Z' - \frac{\tau_a}{\tau_d}\right)^2 + 4\frac{\tau_a}{\tau_\star}Z' \right]^{1/2} \right\}~,
\end{equation}
where $\tau_d$ is the effective dust destruction time scale; $Z'$ is the metallicity relative to solar value. Again, we use $\metal_\odot=8.69$~\footnote{Note that \citet{ANIANO20} use $\metal_\odot=8.75$.}; $f_m$ is the mass fraction of dust-forming metals to mass of total metals, that is:
\begin{equation}
    f_m = \frac{\rm Mass~of~dust{\text -}forming~metals}{\rm Mass~of~total~metals}~.
\end{equation}
In \citet{ANIANO20}, $f_m$ is fixed at $\sim 45.5\%$. The dust injection time scale has minor impact on the prediction. We fix $\tau_a/\tau_\star=10^{-2}$ through out all models since we expect $\tau_a \sim \mathcal{O}(10^7)$\,yr and $\tau_\star \sim \mathcal{O}(10^9)$\,yr in the nearby spiral galaxies, e.g., discussions in \citet{Draine09} and \citet{ASANO13}. Note that we do not expect our measurements to follow one single parameter set because the variation in gas density and SFR will reflect on the change in $\tau_a/\tau_d$ \citep{ASANO13,ANIANO20}.

\begin{figure*}[!ht]
\centerline{\includegraphics[width=0.95\textwidth]{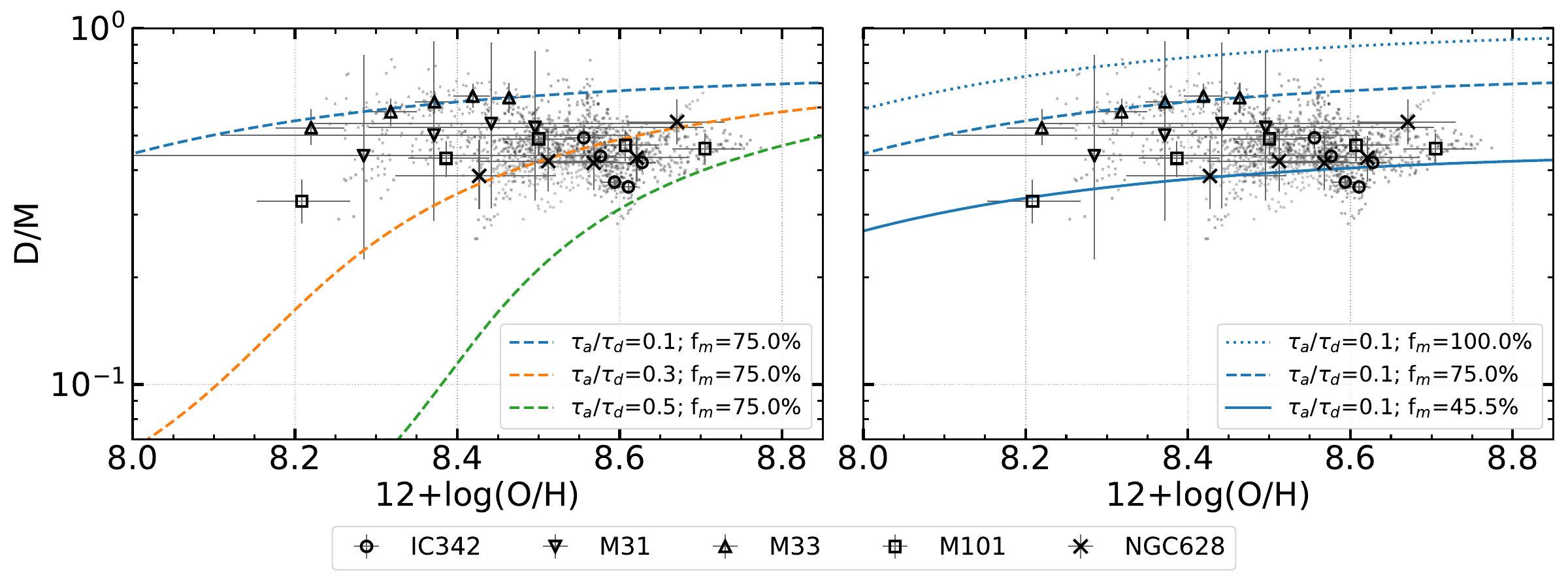}}
\caption{Our measurements and the \citet{ANIANO20} dust evolution model. Left: Fixing $f_m$ and varying $\tau_a/\tau_d$, showing that the variation of D/M is smaller with lower $\tau_a/\tau_d$. Right: Fixing $\tau_a/\tau_d$ and varying $f_m$, showing that larger $f_m$ results in higher D/M. Black: This work. Each marker corresponds to each galaxy as shown in the legend.\label{fig: aniano+20}}
\end{figure*}

We overlay our observed D/M with the model predictions in Figure~\ref{fig: aniano+20}. All the models predict a higher D/M at higher \metal, and the D/M asymptotically approaches $f_m$ toward high \metal. In the left panel, we fix $f_m=75\%$ and plot three different $\tau_a/\tau_d$ ratios: 0.1, 0.3 and 0.5. As dust growth becomes faster relative to dust destruction ($\tau_a/\tau_d$ decreases), the model predicts a smaller variance in D/M in our observed \metal range. In other words, with a lower $\tau_a/\tau_d$, D/M approaches $f_m$ at a smaller \metal.

In the right panel of Figure~\ref{fig: aniano+20}, we fix $\tau_a/\tau_d$ at 0.1 and vary $f_m$. The major part of our measurements have D/M above the $f_m=45.5\%$ model, which means that the fraction of dust-forming metals is probably higher than the value estimated in \citet{ANIANO20}. For IC342, M31, M101, and NGC628, most measured D/M values are between the $f_m=45.5\%$ and $f_m=75\%$ models. This could indicate that the fraction of dust-forming metals is lower than 75\% in these galaxies. M33 has the overall highest D/M. Within the frame of this model, it can indicate that the chemical composition of the ISM or dust is different in M33, or $\tau_a/\tau_d$ in M33 is smaller than in the other galaxies. We will discuss more in Sect.~\ref{sec: discussions-special cases}.

\subsection{Previous Multi-Galaxy Observations of D/M}

\begin{figure*}[!ht]
\centerline{\includegraphics[width=\textwidth]{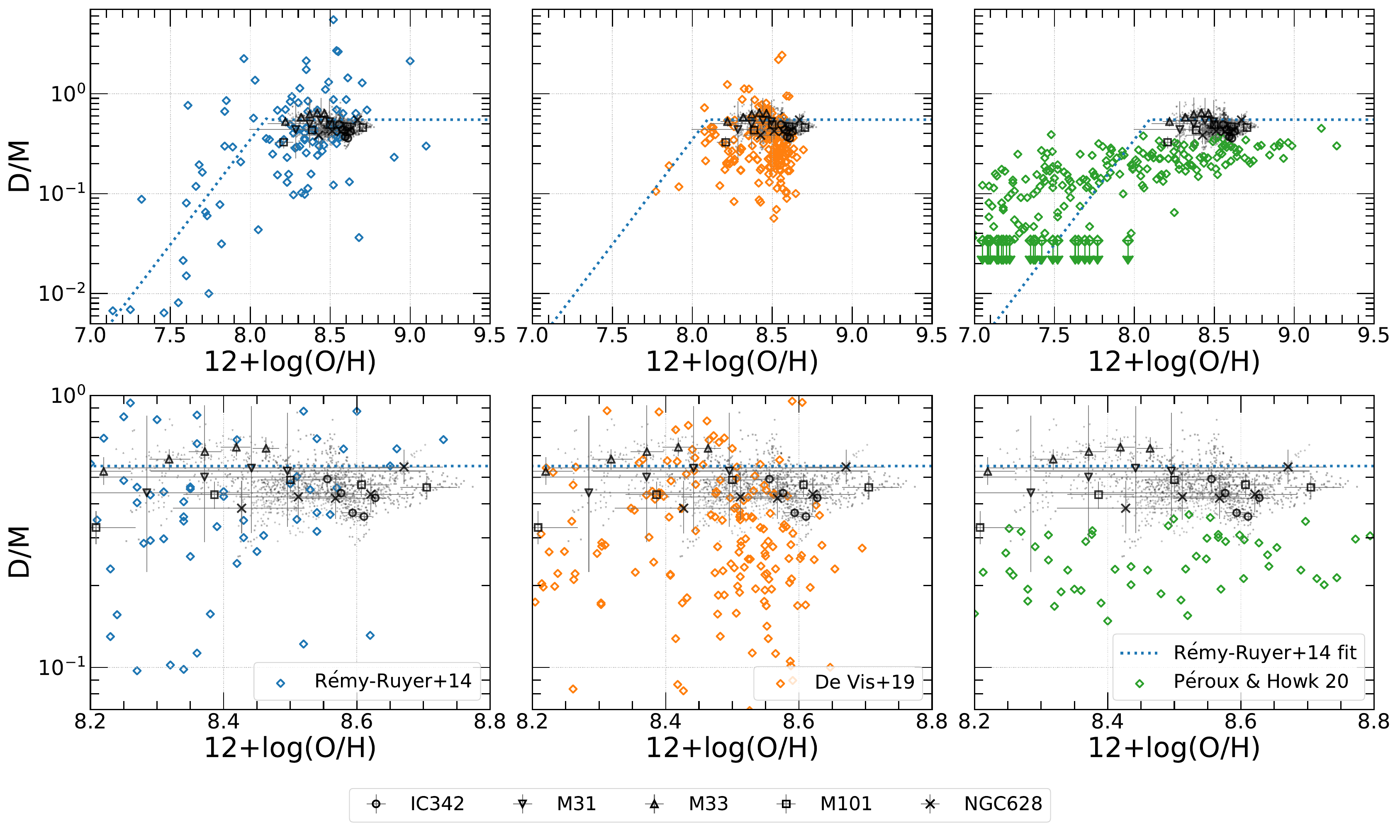}}
\caption{D/M measured in this work overlay with previous observations. Data points with arrows shows the upper limit. Dotted line: the broken power-law fit from \citet[][table~1]{REMY-RUYER14}. Black: This work. Each marker corresponds to each galaxy as shown in the legend. The bottom panels are a zoom-in of the top panels. Left: the \citet{REMY-RUYER14} observations ($\rm X_{CO,\,Z}$ case). Middle: the \citet{DEVIS19} observations (PG16S calibration). Right: the \citet{PerouxHowk20} observations. Our measurements are consistent with \citet{REMY-RUYER14} and \citet{DEVIS19}, while the \citet{PerouxHowk20} measurements are systematically lower than ours. \label{fig: prev obs}}
\end{figure*}

\begin{figure*}[!ht]
\centerline{\includegraphics[width=\textwidth]{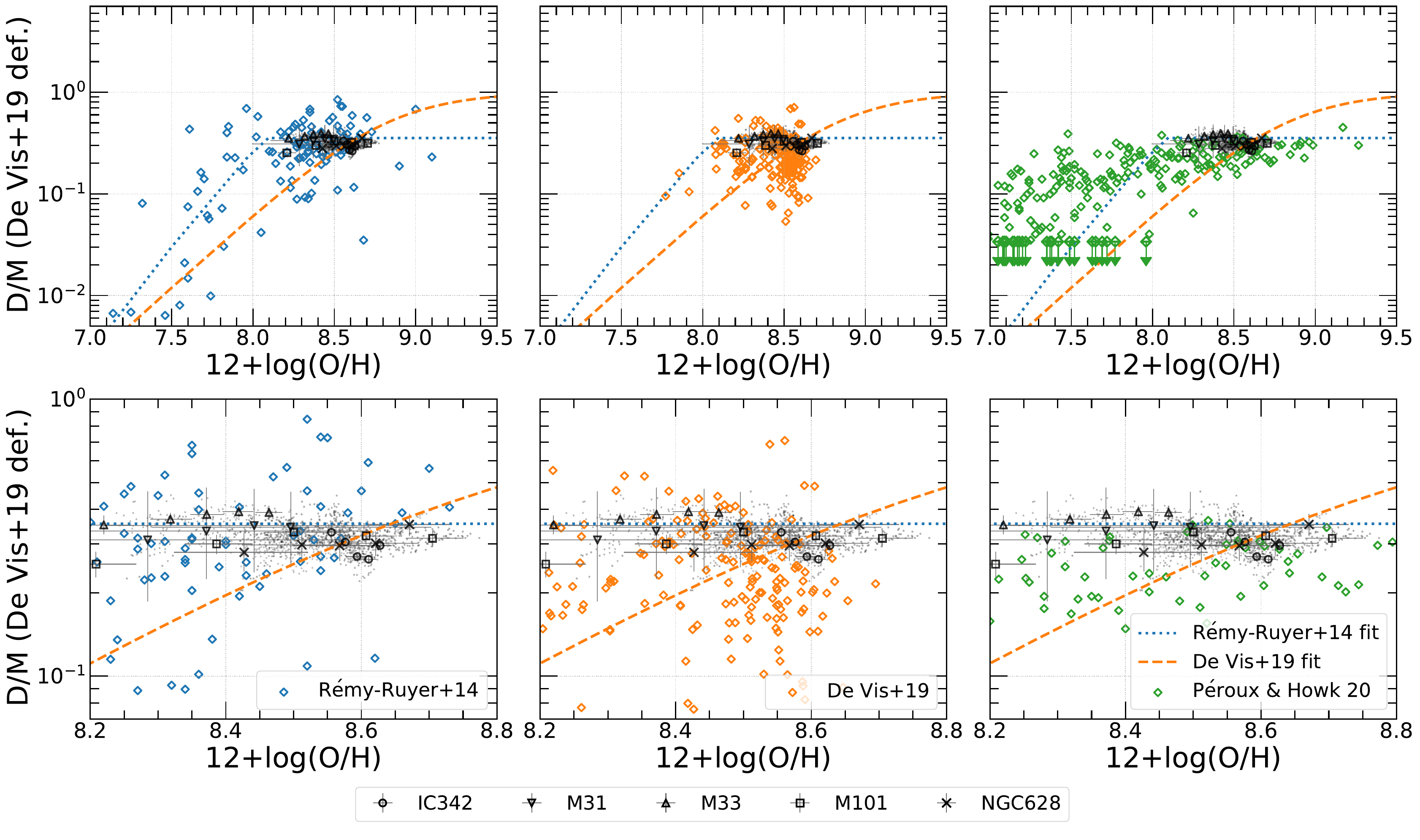}}
\caption{Same as Figure~\ref{fig: prev obs}, but plotted with the D/M definition in \citet{DEVIS19} (except the \citet{PerouxHowk20} data, see main text). Orange dashed line: the \citet{DEVIS19} D/G-\metal fit, converted to D/M-\metal space.\label{fig: prev obs D19}}
\end{figure*}

In this section, we compare our measured D/M to previous multi-galaxy D/M observations, including both IR-based and abundance-based measurements. Further, we inspect if there are significant differences in the D/M-metallicity relations between the resolved and galaxy-integrated measurements.

\citet{REMY-RUYER14} and \citet{DEVIS19} are two IR-based galaxy-integrated studies in the nearby universe. With sample size $>100$ galaxies, both works are the benchmarks of our current understanding of the galaxy-integrated dust properties in the nearby universe. \citet{PerouxHowk20} derive D/M from elemental abundance ratio with the dust-correction model \citep{DeCia16,DeCia18} in $\gtrsim 200$ Damped Lyman-$\alpha$ systems. Their samples have redshifts ranging $0.1 \lesssim z \lesssim 5$, providing us a point of view with different sample selection and methodology. We quote \citet{REMY-RUYER14} data points from their \mbox{table~A.1}; \citet{DEVIS19} have their data public on their website\footnote{\url{http://dustpedia.astro.noa.gr/}}; \citet{PerouxHowk20} include the data table as one of their supplement materials.

Since \citet{DEVIS19} adopt a different definition of D/M from ours, we show the comparison with both definitions. In this work, we assume that the depletion in \HII regions, where we get the \metal measurements, is negligible \citep[$\lesssim 0.1\,\dex$, e.g.][]{Peimbert10}; on the other hand, \citet{DEVIS19} assume the \metal measured in \HII regions only traces gas-phase metals, thus one needs to consider the mass locked in dust grains to get the total metal mass\footnote{In an environment with $\rm D/M \sim 0.5$ in our definition, the ``dust correction'' to metallicity under \citet{DEVIS19} definition is effectively $\sim +0.18\,\dex$, which seems to be overestimating the available metals compared to the estimated depletion in \HII regions.}, that is:
\begin{equation}
    \rm {(D/M)}_{D19} \equiv \frac{\Sigmad}{\Sigmad + \Sigmametal}~.
\end{equation}
We show the D/M derived with our definition in Figure~\ref{fig: prev obs}, and the D/M derived with the \citet{DEVIS19} definition in Figure~\ref{fig: prev obs D19}. Note that the \citet{PerouxHowk20} measurements are not converted in either figure due to their D/M-derivation methodology. \citet{PerouxHowk20} derive D/M with a dust-corrected model \citep{DeCia16,DeCia18}, which already includes both the gas-phase metal and metal in dust.

In Figures \ref{fig: prev obs}~and~\ref{fig: prev obs D19}, we show the \citet{REMY-RUYER14} measurements in the left panels and their fit (the gas-to-dust ratio fit with a broken power-law converted to D/M-to-\metal. $\rm X_{CO,\,Z}$ case) in all panels. Our measurements locate roughly in the center of their measurements at the same metallicity, and our measurements are also consistent with their broken power law in the high-metallicity end. Both facts suggest that our D/M-to-metallicity relations are consistent with the \citet{REMY-RUYER14} measurements. Note that in the high-metallicity region, the \citet{REMY-RUYER14} broken power law gives a constant D/M, which matches our measurements that D/M is roughly a constant across galaxies. In Figure~\ref{fig: prev obs}, there are some \citet{REMY-RUYER14} measurements with $\text{D/M}>1$. Since their adopted \aco \citep{SCHRUBA12} has relatively large normalization (see Figure~\ref{fig: MetalDensity}), those high D/M values are not likely due to underestimating \Sigmagas from the choice of \aco. Instead, it is more likely an issue in the adopted dust opacity function, differences in dust SED fitting techniques, or differences in metallicity calibration.

In the middle panels of Figures \ref{fig: prev obs}~and~\ref{fig: prev obs D19}, we present the \citet{DEVIS19} measurements (PG16S calibration) and ours. We only select data points where both \HI and \HTWO measurements are available in \citet{DEVIS19}. We also convert their D/G-\metal fit to a D/M-\metal and plot it in all panels in Figure~\ref{fig: prev obs D19}. Note that this fit is not created for the purpose of predicting D/M, thus it is possible to generate unphysical D/M at high metallicity due to its power-law nature in our definition of D/M. Our measured D/M scatters around the upper end of the \citet{DEVIS19} data range in both figures.

We present the \citet{PerouxHowk20} measurements in the right panels of Figures \ref{fig: prev obs}~and~\ref{fig: prev obs D19}. \citet{PerouxHowk20} derived their D/M by converting observed elemental abundance ratios into depletion with the empirical formulas in \citet{DeCia16,DeCia18}. They show that as metallicity increases, D/M increases and the scatter of D/M decreases. This trend is shown over all redshifts. In Figure~\ref{fig: prev obs}, our measured D/M is systematically higher than the D/M in \citet{PerouxHowk20}. There are two possible causes for the offset. First, the sample selection in \citet{PerouxHowk20} is based on \HI column density and a lot of the data comes from \HI-dominated regions, while most of our data points are in \HTWO-dominated regions. In other words, the offset might come from the difference in dust evolution in \HI- and \HTWO-dominated regions. Second, there might be a systematic offset between the IR-based and abundance-based D/M determination.

\subsection{The High D/M in M33}\label{sec: discussions-special cases}
In our measurements, we find M33 has a higher D/M than the other galaxies at the same metallicity. One possibility is that the \aco in M33 is larger than \acoBolatto. In \citet{GRATIER10} and \citet{DRUARD14}, the authors suggest a constant $\aco=2\acoMW$, which is larger than \acoBolatto everywhere in M33. If we use 2\acoMW for M33, the median D/M in M33 will slightly decrease from 0.60 to 0.56, which brings it closer to the other galaxies.

On the other hand, we could also try to interpret this higher D/M with the \citet{ANIANO20} dust evolution model. The first possibility is that $f_m$ is higher in M33. That means the ISM chemical composition is different in M33, and there is a larger fraction of dust-forming metals, or a higher ratio of dust-forming metals to oxygen abundance. The second possibility is a shorter $\tau_a$ or a longer $\tau_d$ in M33. This explanation is less likely because M33 does not seem to have a higher \PDE or \Sigmasfr relative to the other galaxies, which are the two key factors affecting $\tau_a$ and $\tau_d$.

\subsection{Future Perspectives in \texorpdfstring{\aco}{the CO-to-H2 Conversion Factor} Constraints}\label{sec: discussions-alpha_CO perspectives}

In Sect.~\ref{sec: results-D/M+aco}, we show that D/M is sensitive to the choice of \aco. We find the most reasonable D/M with \acoBolatto, however, we still have negative D/M-metallicity and D/M-density correlations, especially in IC342. In Sect.~\ref{sec: results-constrain}, we demonstrate that a \aco prescription described by simple power law with metallicity is not enough to solve the negative correlations. We need a more complex functional form, or taking the effects from other environmental parameters into account, e.g.\ gas temperature or velocity dispersion. Unfortunately, we do not have enough data points with high \Sigmatot, and currently the fitting results are biased toward the centers of IC342 if we do adopt the \citet{BOLATTO13} functional form with our constraints.

To continue investigating on the effect of \Sigmatot, one needs to study nearby galaxies with high resolution \Sigmagas data, e.g., the PHANGS (The Physics at High Angular resolution in Nearby GalaxieS Surveys, A.~K.\ Leroy et al.\ in preparation) galaxies. Meanwhile, the analysis is also limited by the resolution of dust maps and FIR observations. Among the existing and retired FIR telescopes, \textit{Herschel} has the highest spatial resolution, $\sim 1.8\,\kpc$ at a distance of 10\,\Mpc. The resolution is not enough if we want to resolve a $< 1\,\kpc$ high surface density region. A future mission of FIR photometry at higher resolution is needed to improve our understanding of ISM dust.

Meanwhile, the \citet{BOLATTO13} functional form is less applicable to distant galaxies because it is built on resolved measurements of \Sigmagas and \Sigmastar. One possible approach to apply the \acoBolatto prescription to distant galaxies is to derive a conversion from galaxy-integrated quantities to total molecular gas mass derived with \acoBolatto in resolved galaxies. A larger sample of galaxies with \CO emission, stellar mass, and resolved metallicity measurements is required for this approach. Auxiliary data like SFR and total atomic gas mass might also be helpful in the derivation.

\section{Summary}\label{sec: summary}
We investigate the relation between dust-to-metals ratio (D/M) and various local ISM environmental quantities in five nearby galaxies: IC342, M31, M33, M101, and NGC628. The multi-wavelength data from both archival and new observations are processed uniformly. 
A modified blackbody model with a broken power-law emissivity is used to model the dust emission SED, together with the fitting techniques and dust opacity calibration proposed by \cite{GORDON14} and implemented in \citet{CHIANG18} (Sect.~\ref{sec: data-dust}). We utilize metallicity gradients derived from auroral line measurements in \HII regions to ensure a uniform and high-quality metallicity determination wherever possible. We calibrate and image a new IC342 \HI 21\,\cm map from new VLA observations. This is part of the observations in the EveryTHINGS project (P.I.\ K.~M.\ Sandstrom; I.\ Chiang et al.\ in preparation). All maps are convolved to a common physical resolution at $\sim2\,\kpc$ for a uniform analysis.

We propose a new approach to constrain D/M and the \CO-to-\HTWO conversion factors (\aco), that is, we use the expected D/M-metallicity and D/M-ISM gas density correlations measured by depletion studies to evaluate the results. We use this conceptual approach to examine the D/M yielded by existing \aco prescriptions, and demonstrate our first attempt in utilizing this approach to constrain simple metallicity power-law \aco. We find the following key points:
\begin{enumerate}
    \item Among the prescriptions we test, \acoBolatto yields the most reasonable D/M.
    \item With \acoBolatto, the D/M is roughly a constant ($0.46^{+0.12}_{-0.06}$) across a large range of ISM environments.
    \item When we exclude IC342, \acoMW and \acoHunt can satisfy most constraints set by the D/M-metallicity and D/M-\PDE correlations, while \acoSchruba seems to have a normalization that is too high (2\acoMW at $Z_\odot$).
    \item The most obvious difference between \acoBolatto and other prescriptions is the dependence on the total surface density ($\Sigmatot=\Sigmagas+\Sigmastar$), which decreases \aco in the regions with $\Sigmatot > 100\,\SigmaMassUnit$. This is mostly important in the centers of galaxies, and likely starburst regions.
    \item To properly account for the \HTWO gas in IC342, it seems that an \aco prescription parameterized by \metal only is not enough. The \acoBolatto, which depends on \Sigmatot, yields the most reasonable D/M in IC342.
    \item New FIR observations with spatial resolution better than \textit{Herschel} are needed for investigating D/M and \aco at high surface density regions.
\end{enumerate}

In Sect.~\ref{sec: discussions}, we interpret our observations with the dust evolution model from \citet{ANIANO20}. We also compare our results to the previous galaxy-integrated D/M measurements. We find the following implications regarding our results:
\begin{enumerate}
    \item The roughly constant D/M implies a shorter dust growth time scale ($\tau_a$) relative to the dust destruction time scale ($\tau_d$).
    \item Most of our measurements fall in the range between $f_m=45.5\%$ and $f_m=75\%$, with $f_m$ being the mass fraction of dust forming metals.
    \item Our measured D/M is consistent with previous IR-based, galaxy-integrated measurements in the nearby universe \citep{REMY-RUYER14,DEVIS19}.
    \item However, our results are systematically higher than the D/M measured in the abundance-based measurements by \citet{PerouxHowk20}. This could indicate a systematic offset between IR-based and abundance-based methods.
\end{enumerate}
Our results demonstrate that D/M is sensitive to the choice of \aco. The \acoBolatto is our current best choice of \aco, which models the decrease of \aco due to gas temperature and velocity dispersion. Our results show a roughly constant D/M across ISM environments. Further investigation is needed to constrain D/M and \aco simultaneously.

\acknowledgments
We thank the referee for thoughtful comments that improved
the paper. We gratefully acknowledge the hard work of the CHAOS, DustPedia, EDD, HERACLES, HerM33s, MaNGA, PHANGS, THINGS, $z$0MGS teams and thank them for making their data publicly available.

IC thanks C.~R.\ Choban, H.~Hirashita, C.~Howk, D.~Kere\v{s} and S.~Zhukovska for the useful discussions about to this work. The work of KS, IC, AKL, DU and JC is supported by National Science Foundation grant No. 1615728 and NASA ADAP grants NNX16AF48G and NNX17AF39G. The work of AKL and DU is partially supported by the National Science Foundation under Grants No. 1615105, 1615109, and 1653300. KK gratefully acknowledges funding from the Deutsche Forschungsgemeinschaft (DFG, German Research Foundation) in the form of an Emmy Noether Research Group (grant number KR4598/2-1). TGW acknowledges funding from the European Research Council (ERC) under the European Union’s Horizon 2020 research and innovation programme (grant agreement No. 694343). JP and CH acknowledge support from the Programme National ``Physique et Chimie du Milieu Interstellaire'' (PCMI) of CNRS/INSU with INC/INP, co-funded by CEA and CNES.

This work uses observations made with ESA \textit{Herschel} Space Observatory. \textit{Herschel} is an ESA space observatory with science instruments provided by European-led Principal Investigator consortia and with important participation from NASA. The \textit{Herschel} spacecraft was designed, built, tested, and launched under a contract to ESA managed by the \textit{Herschel}/Planck Project team by an industrial consortium under the overall responsibility of the prime contractor Thales Alenia Space (Cannes), and including Astrium (Friedrichshafen) responsible for the payload module and for system testing at spacecraft level, Thales Alenia Space (Turin) responsible for the service module, and Astrium (Toulouse) responsible for the telescope, with in excess of a hundred subcontractors.

The National Radio Astronomy Observatory is a facility of the National Science Foundation operated under cooperative agreement by Associated Universities, Inc. This research is based on observations made with the Galaxy Evolution Explorer (GALEX), obtained from the MAST data archive at the Space Telescope Science Institute, which is operated by the Association of Universities for Research in Astronomy, Inc., under NASA contract NAS 5–26555. This work is based on observations carried out with the IRAM NOEMA Interferometer and 30m telescope. IRAM is supported by INSU/CNRS (France), MPG (Germany) and IGN (Spain). This publication makes use of data products from the Wide-field Infrared Survey Explorer, which is a joint project of the University of California, Los Angeles, and the Jet Propulsion Laboratory/California Institute of Technology, funded by the National Aeronautics and Space Administration. The WSRT is operated by ASTRON (Netherlands Foundation for Research in Astronomy) with support from the Netherlands Foundation for Scientific Research (NWO).

This research made use of Astropy,\footnote{http://www.astropy.org} a community-developed core Python package for Astronomy \citep{ASTROPY13,Astropy18}. This research has made use of NASA's Astrophysics Data System Bibliographic Services. We acknowledge the usage of the HyperLeda database (http://leda.univ-lyon1.fr). This research has made use of the NASA/IPAC Extragalactic Database (NED), which is funded by the National Aeronautics and Space Administration and operated by the California Institute of Technology.

\vspace{5mm}
\facilities{GALEX, \textit{Herschel}, IRAM:30m, IRAM:NOEMA, VLA, WISE, WSRT}
\software{Astropy \citep{ASTROPY13,Astropy18},
          CASA \citep{MCMULLIN07},
          GILDAS\footnote{http://www.iram.fr/IRAMFR/GILDAS},
          Matplotlib \citep{HUNTER07},
          NumPy \& SciPy \citep{VANDERWALT11},
          pandas \citep{MCKINNEY10},
          SAOImage DS9 \citep{JOYE03}
          }

\appendix
\section{Measurements with \texorpdfstring{\acoBolatto}{Bolatto et al. (2013) Conversion Factor}}\label{app: results-other}
\begin{figure*}[!ht]
\centerline{\includegraphics[width=\textwidth]{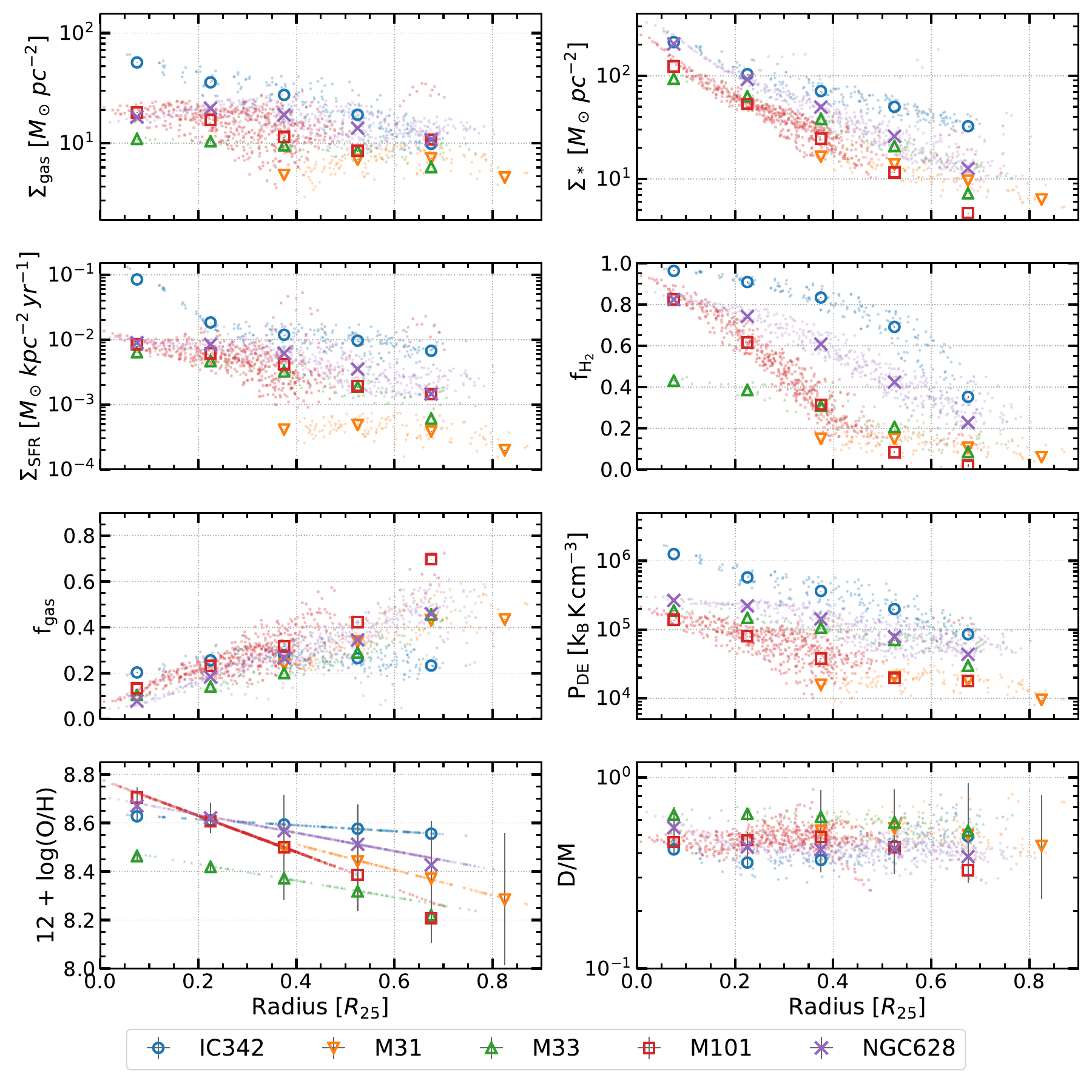}}
\caption{Radial profiles of the observed and derived quantities calculated with \acoBolatto. Small markers: pixel-by-pixel data where detection is above $3\sigma$. Large markers: average in radial bins. The errorbar shows the 16-/84-percentile distribution from 1000 Monte Carlo tests, assuming Gaussian error in measurements.\label{fig: profile B13}}
\end{figure*}

\begin{figure*}[!ht]
\centerline{\includegraphics[width=\textwidth]{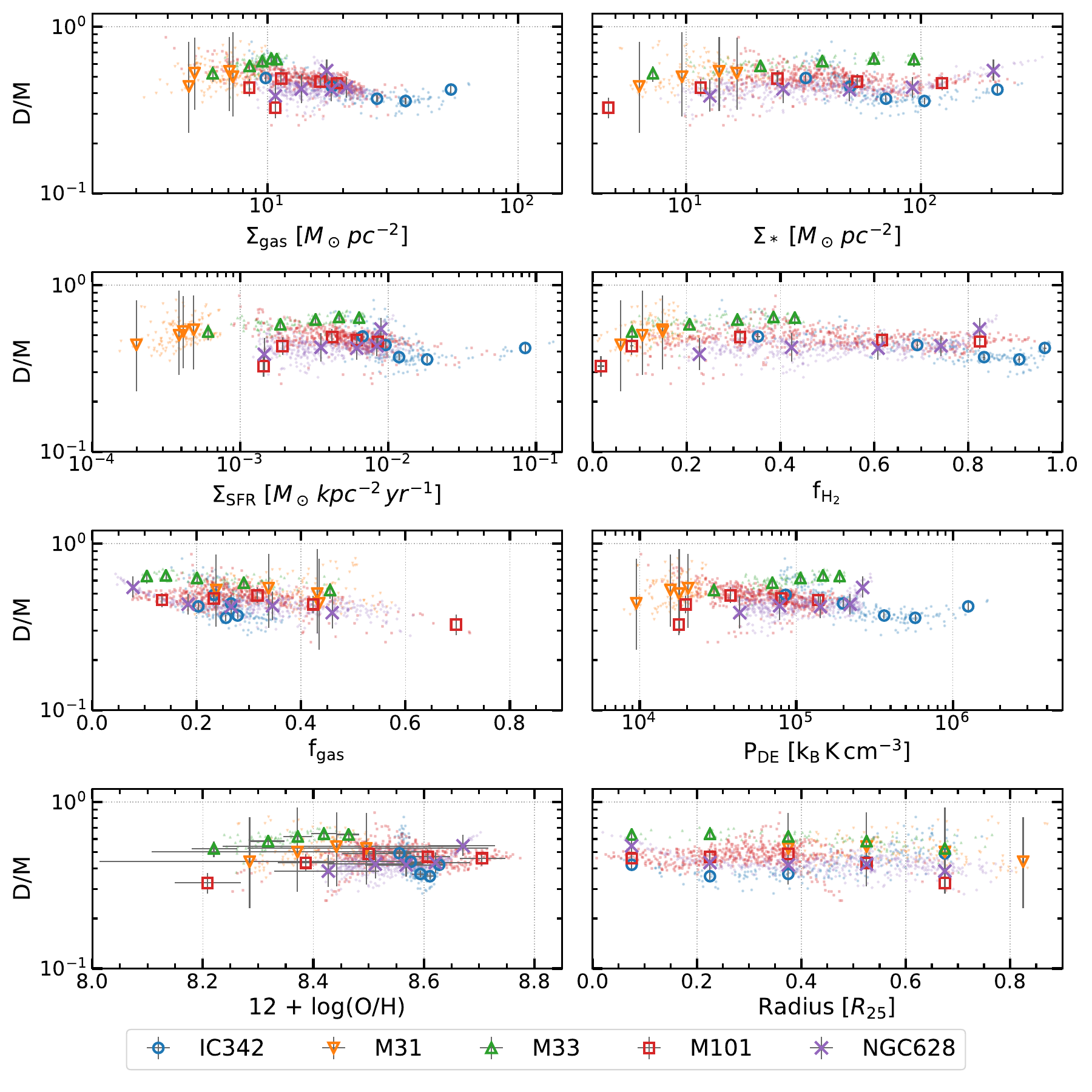}}
\caption{Relations between the physical quantities and D/M  calculated with \acoBolatto. The variation of D/M is minimum across all quantities displayed. Small markers: pixel-by-pixel data where detection is above $3\sigma$. Large markers: average in radial bins. The errorbar shows the 16-/84-percentile distribution from 1000 Monte Carlo tests, assuming Gaussian error in measurements.\label{fig: D/M-ISM B13}}
\end{figure*}

As we have stated in Sect.~\ref{sec: results-summary}, due to our limited understanding of the \CO-to-\HTWO conversion factor, we do not intend to conclusively determine the environmental dependence of D/M with the current measurements. However, it is still informative to show our measurements with \acoBolatto here. We present the radial profiles of the measured and derived quantities in Figure~\ref{fig: profile B13}. D/M is roughly constant as radius increases. On the other hand, most measured quantities decreases as radius increases except \fgas, which increases with radius. 

Figure~\ref{fig: D/M-ISM B13} shows the relationship between the physical quantities and D/M. Generally speaking, the D/M is roughly constant across most physical environments. We also notice D/M tend to decrease as \fgas increases, which is a similar trend found in the galaxy-integrated measurements in \citet{DEVIS19}.

\bibliographystyle{yahapj}
\bibliography{idchiang}

\end{document}

%% file: tab_alldata.tex
\begin{deluxetable}{llllll}
\tablewidth{0pt} 
\tablecaption{Properties of selected galaxies\tablenotemark{a}}\label{tab: data}
\tablehead{
\colhead{Name}                  & \colhead{Morph.}          & \colhead{Distance\tablenotemark{b}}             &
\colhead{Incl.}                 & \colhead{P.A.}            & \colhead{\R{25}}            \\
                                &                           & \colhead{[\Mpc]}               &
\colhead{[degrees]}             & \colhead{[degrees]}       & \colhead{[\arcmin]}           
}
\startdata
IC342                           & SABc                      & 3.45\tablenotemark{c}                          &
18.46                           & \nodata\tablenotemark{d}                   & 9.88                          \\
M31                             & Sb                        & 0.79                          &
77.7\tablenotemark{e}                      & 38.0\tablenotemark{e}                & 88.9                          \\
M33                             & Sc                        & 0.92                          &
55.0\tablenotemark{f}                      & 200.0\tablenotemark{f}               & 31.0                          \\
M101                            & SABc                      & 6.96                          &
18.0\tablenotemark{g}                      & 39.0\tablenotemark{g}                & 12.0                          \\
NGC628                          & Sc                        & 9.77\tablenotemark{h}                    &
8.7\tablenotemark{i}                      & 20.8\tablenotemark{i}                & 4.94                          \\
\enddata
\vspace{1ex}
\tablerefs{\tablenotemark{a} The HyperLeda database \citep{MAKAROV14}. \tablenotemark{b} The Extragalactic Distance Database \citep[EDD,][]{TULLY09}. \tablenotemark{c} \citet{WU14}. \tablenotemark{d} Treated as $0.0^\circ$ because it is a face-on galaxy. \tablenotemark{e} \citet{CORBELLI10}. \tablenotemark{f} \citet{KOCH18}. \tablenotemark{g} \citet{SOFUE99}. \tablenotemark{h} \citet{MCQUINN17}. \tablenotemark{i} \citet{Lang20}.}
\end{deluxetable}

%% file: tab_aco_prescriptions.tex
\renewcommand{\arraystretch}{1.5}
\begin{deluxetable}{lc}
\tablewidth{0pt} 
\tablecaption{List of $\aco$ prescriptions used in this work.\label{tab: aco_prescription}}
\tablehead{
\colhead{Prescription}                  &
\colhead{$\aco$ formula}                \\
\colhead{}                              &
\colhead{[\acoUnit]}                    
}
\startdata
\acoMW                                  &
4.35                                    \\
\acoSchruba                             &
$8.0 \times\rm (Z/Z_\odot)^{~-2.0}$     \\
\acoBolatto                             &
$\rm 2.9\times\exp\left(\frac{0.4}{Z/Z_\odot}\right)\times\left\{\begin{array}{ll}
    \left(\Sigma_{\rm Total}^{100}\right)^{-0.5} & ,\,\Sigma_{\rm Total}^{100} \geq 1 \\
    1 & ,\,\Sigma_{\rm Total}^{100} < 1
\end{array}\right.$                     \\
\acoHunt                                &
$\rm 4.35\times\left\{\begin{array}{ll}
    1 & ,\,Z \geq Z_\odot \\
    (Z/Z_\odot)^{-1.96} & ,\,Z < Z_\odot
\end{array}\right.$                     \\
\enddata
\tablecomments{$\Sigma_{\rm Total}^{100}$ is \Sigmatot in $100\,\SigmaMassUnit$. $Z/Z_\odot=1$ at $\metal=8.69$ throughout this paper.}
\end{deluxetable}
\renewcommand{\arraystretch}{1.0}

%% file: tab_metaldata.tex
\begin{deluxetable*}{lcccc}[!tb]
\tablewidth{0pt} 
\tablecaption{\metal data.\label{tab: metal}}
\tablehead{
\colhead{Name}          &
\colhead{Reference}     &
\colhead{\metal at the galaxy center}    &
\multicolumn{2}{c}{Slope\tablenotemark{a}} \\
                                &
                                &
\colhead{[dex]}                 &
\colhead{[$\dex\,\kpc^{-1}$]}      &
\colhead{[$\dex\,\R{25}^{-1}$]}   
}
\startdata
IC342                   &
K.~Kreckel et al.\ in preparation\tablenotemark{b} &
8.64 ($\pm0.01$)        &
-0.012 ($\pm 0.003$)     &
-0.12 ($\pm 0.03$)       \\
M31                     &
\citet{ZURITA12}        &
8.72 ($\pm$0.18)        &
-0.026 ($\pm$0.013)      &
-0.52 ($\pm$0.26)        \\
M33                     &
\citet{BRESOLIN11_M33}  &
8.50 ($\pm$0.02)        &
-0.041 ($\pm$0.005)      &
-0.34 ($\pm$0.04)        \\
M101                    &
\citet{CROXALL16,Berg20}&
8.78 ($\pm$0.04)        &
-0.031 ($\pm$0.002)      &
-0.75 ($\pm$0.06)        \\
NGC628                  &
\citet{BERG15,Berg20}   &
8.71 ($\pm$0.06)        &
-0.027 ($\pm$0.007)      &
-0.38 ($\pm$0.10)        \\
\enddata
\tablenotetext{a}{The slopes have been converted to account for the distances and \R{25} values adapted in this paper.}
\tablenotetext{b}{Using the S-calibration from \citet{PilyuginGrebel16}.}
\end{deluxetable*}

%% file: tab_metaldensity_corr_MonteCarloRadial.tex
\begin{deluxetable*}{llcccc}[!tb]
\tablewidth{0pt} 
\tablecaption{Pearson correlation coefficients of the radial D/M-\metal and D/M-\PDE dependence within each galaxy. The upper and lower variances are quoted from the 16- and 84-percentiles of 1000 Monte Carlo tests (see text).\label{tab: metaldensity_corr}}
\tablehead{
\multirow{2}{*}{Galaxy}           &
\multirow{2}{*}{Quantity}         &
\multicolumn{4}{c}{Prescription}  \\
              &
              &
\acoMW        &
\acoSchruba   &
\acoBolatto   &
\acoHunt
}
\startdata
\multirow{2}{*}{IC342}  & corr(D/M, \metal)    &
$-0.99^{+0.02}_{-0.01}$ & $-0.99^{+0.01}_{-0.01}$ & $-0.65^{+0.23}_{-0.13}$ & $-0.99^{+0.01}_{-0.01}$ \\
                        & corr(D/M, \PDE)      &
$-0.89^{+0.01}_{-0.10}$ & $-0.78^{+0.01}_{-0.21}$ & $-0.31^{+0.01}_{-0.46}$ & $-0.83^{+0.01}_{-0.16}$  \\
\hline
\multirow{2}{*}{M31}    & corr(D/M, \metal)    &
$0.90^{+0.09}_{-0.99}$ & $0.99^{+0.01}_{-1.79}$ & $0.92^{+0.07}_{-1.07}$ & $0.98^{+0.01}_{-1.25}$ \\
                        & corr(D/M, \PDE)      &
$0.93^{+0.01}_{-0.36}$ & $0.54^{+0.21}_{-1.01}$ & $0.91^{+0.02}_{-0.38}$ & $0.75^{+0.09}_{-0.49}$  \\
\hline
\multirow{2}{*}{M33}    & corr(D/M, \metal)    &
$0.94^{+0.02}_{-0.28}$ & $-0.89^{+0.35}_{-0.05}$ & $0.97^{+0.01}_{-0.43}$ & $-0.02^{+0.71}_{-0.60}$ \\
                        & corr(D/M, \PDE)      &
$0.86^{+0.11}_{-0.18}$ & $-0.81^{+0.17}_{-0.14}$ & $0.91^{+0.05}_{-0.34}$ & $0.13^{+0.52}_{-0.78}$ \\
\hline
\multirow{2}{*}{M101}   & corr(D/M, \metal)    &
$0.42^{+0.17}_{-0.14}$ & $-0.83^{+0.21}_{-0.08}$ & $0.82^{+0.07}_{-0.08}$ & $0.75^{+0.14}_{-0.20}$ \\
                        & corr(D/M, \PDE)      &
$-0.13^{+0.42}_{-0.01}$ & $-0.89^{+0.12}_{-0.08}$ & $0.47^{+0.25}_{-0.01}$ & $0.40^{+0.34}_{-0.06}$ \\
\hline
\multirow{2}{*}{NGC628} & corr(D/M, \metal)    &
$-0.30^{+0.97}_{-0.43}$ & $-0.63^{+1.40}_{-0.27}$ & $0.85^{+0.09}_{-0.18}$ & $0.96^{+0.03}_{-0.29}$ \\
                        & corr(D/M, \PDE)      &
$-0.62^{+1.22}_{-0.17}$ & $-0.46^{+1.18}_{-0.47}$ & $0.84^{+0.05}_{-0.26}$ & $0.98^{+0.01}_{-0.35}$
\enddata
\end{deluxetable*}